\documentclass[10pt,prd,aps,amssymb,amsmath,tightenlines,twocolumn]{revtex4} 

\usepackage{graphicx,epsfig,xcolor}
\usepackage[T1]{fontenc}
\usepackage[latin1]{inputenc}
\usepackage[english]{babel}
\usepackage{amsmath}
\usepackage{mathtools}
\usepackage{amsfonts}
\usepackage{amssymb}
\usepackage{dsfont}

\newcommand{\cL}{\ensuremath{\mathcal{L}}}

\newcommand{\cD}{\ensuremath{\mathcal{D}}}

\newcommand{\cPT}{\ensuremath{\mathcal{PT}}}

\newcommand{\half}{\mbox{$\textstyle{\frac{1}{2}}$}}

\newcommand{\threehalf}{\mbox{$\textstyle{\frac{3}{2}}$}}
\newcommand{\threefourth}{\mbox{$\textstyle{\frac{3}{4}}$}}
\newcommand{\third}{\mbox{$\textstyle{\frac{1}{3}}$}}
\newcommand{\tthird}{\mbox{$\textstyle{\frac{2}{3}}$}}
\newcommand{\fourth}{\mbox{$\textstyle{\frac{1}{4}}$}}

\newcommand{\tfth}{\mbox{$\textstyle{\frac{2}{15}}$}}

\begin{document}

\title{Dyson-Schwinger equations in zero dimensions and polynomial
approximations}
\author{Carl M.~Bender$^a$}\email{cmb@wustl.edu}
\author{C.~Karapoulitidis$^b$}
\email{christos.karapoulitidis@stud.uni-heidelberg.de}
\author{S.~P.~Klevansky$^b$}\email{spk@physik.uni-heidelberg.de}

\affiliation{$^a$Department of Physics, Washington University, St.~Louis,
Missouri 63130, USA\\
$^b$Institut f\"ur Theoretische Physik, Universit\"at Heidelberg, 69120
Heidelberg, Germany\\}

\begin{abstract}
The Dyson-Schwinger (DS) equations for a quantum field theory in $D$-dimensional
space-time are an infinite sequence of coupled integro-differential equations
that are satisfied exactly by the Green's functions of the field theory. This
sequence of equations is {\it underdetermined} because if the infinite sequence
of DS equations is truncated to a finite sequence, there are always more Green's
functions than equations. An approach to this problem is to close the finite
system by setting the highest Green's function(s) to zero. One can examine the
accuracy of this procedure in $D=0$ because in this special case the DS
equations are just a sequence of coupled {\it polynomial} equations whose roots
are the Green's functions. For the closed system one can calculate the roots and
compare them with the exact values of the Green's functions. This procedure
raises a general mathematical question: When do the roots of a sequence of
polynomial approximants to a function converge to the exact roots of that
function? Some roots of the polynomial approximants may (i) converge to the
exact roots of the function, or (ii) approach the exact roots at first and then
veer away, or (iii) converge to limiting values that are unequal to the exact
roots. In this study five field-theory models in $D=0$ are examined, Hermitian
$\phi^4$ and $\phi^6$ theories and non-Hermitian $i\phi^3$, $-\phi^4$, and $-i
\phi^5$ theories. In all cases the sequences of roots converge to limits that
differ by a few percent from the exact answers. Sophisticated asymptotic
techniques are devised that increase the accuracy to one part in $10^7$. Part of
this work appears in abbreviated form in Phys.~Rev.~Lett.~{\bf 130}, 101602
(2023).

% \vskip 0.2cm
% \today
\end{abstract}
\maketitle

\section{Introduction}\label{s1}
In a recent Letter \cite{R1} we examined the effectiveness of the
Dyson-Schwinger (DS) equations to calculate the Green's functions for both
Hermitian and $\cPT$-symmetric quantum field theories. This letter presents in
compact form our studies of five zero-dimensional models: Hermitian $\phi^4$
and $\phi^6$ and non-Hermitian $i\phi^3$, $-\phi^4$, and $-i\phi^5$ theories.
Field theories in $D=0$ are useful because the Green's functions are already
known exactly and the DS equations are polynomial equations in the Green's
functions, so one can evaluate the accuracy of the truncation scheme used to
close the infinite system of coupled DS equations. The current paper presents
the detailed results of this study \cite{R2}.
 
The advantage of studying zero-dimensional field theory is that we can reduce a
very difficult problem -- that of solving the DS equations for the Green's
functions of a field theory -- to the {\it generic} problem of finding the roots
of a polynomial equation. The polynomial depends on the choice of field theory
and also on the scheme that is used to solve the infinite tower of DS equations.
To construct this polynomial we first truncate the infinite sequence of DS
equations to a finite set consisting of the first $N$ coupled polynomial
equations. This finite system is {\it underdetermined} because there are always
more Green's functions than equations. Next, we set all but the first $N$
Green's functions to zero and solve the resulting {\it determined} coupled
polynomial system. This polynomial system is triangular so it is easy to
eliminate successively all but the lowest Green's function, which then satisfies
the $N$th degree polynomial equation $P_N(x)=0$.

This kind of iterative approach in which we take more and more DS equations is
common in field theory: One begins with a {\it leading} approximation and then
constructs a sequence of approximations that one hopes will approach the exact
answer. If we knew the underlying function that the sequence of polynomial
approximants $P_N(x)$ represents, we could use standard techniques such as
Newton's method to determine the roots. However, for difficult problems in
physics, as is the case here, the polynomial $P_N(x)$ is only an approximate
consequence of the DS equations.

We are led to ask, Do the roots of the polynomial approximation at each order
lead to the correct solution, and what is the nature of the convergence (if it
exists)? There are several possibilities: (i) The accuracy of the roots of the
polynomial approximation $P_N(x)$ improve as $N\to\infty$ (that is, as one
includes more DS equations), and some or all of the roots converge to the
correct answer; (ii) The roots of $P_N(x)$ at first approach the correct answer,
but then diverge away from it. The former behavior is characteristic of Taylor
expansions, where, if the sequence of approximants converges, it converges to
the right answer. The latter behavior is characteristic of asymptotic series.
Both (i) and (ii) reflect the usual behavior of series approximations. 

There is also a third possibility: (iii) The roots of $P_N(x)$ converge as
$N\to\infty$, but they converge to the {\it wrong} answer, that is, to a number
that may be {\it close} to the exact answer but is {\it not} the correct answer.
This means that the procedure may be used to gain an approximate understanding
of the physics but that the accuracy of the result is limited. It is rather
unusual for a sequence of approximants to behave in this way.

Our expectation in solving the $D=0$ field-theoretic models for the Green's
functions was that increasing the number of DS equations would lead to {\it
increasing} accuracy in our results if we use the {\it unbiased} procedure of
truncating the DS equations by setting higher Green's functions to zero rather
than guessing the behavior of the higher Green's functions. However, this is
{\it not} the case: The unbiased truncation procedure does {\it not} lead to
convergence to the correct value for the Green's function as we go to higher
orders. Rather, we observe the third possibility (iii). This discovery holds for
both Hermitian and non-Hermitian theories. The only truncation strategy that
appears to work (for both kinds of theories) is to find the {\it asymptotic
behavior} of the Green's function in the limit of large order of truncation;
that is, to find the asymptotic behavior of the $n$th Green's function for large
$n$. Finding this asymptotic behavior is nontrivial. However, if this is done,
we find that order-by-order in the asymptotic approximation, the roots of the
polynomials rapidly get closer to the exact values of the Green's functions.
 
One objective of our study was to search for differences in the convergence
behavior of the DS equations for Hermitian and non-Hermitian field theories.
There are subtle differences in the convergence behavior: Hermitian theories
display a monotone behavior while non-Hermitian theories have an oscillatory
behavior.
 
This paper is organized as follows. In Sec.~\ref{s2} we use a parabolic cylinder
function to illustrate the difficulties with calculating the zeros of a function
from polynomial approximations to that function. The question is whether the
sequence of polynomials obtained from a Taylor series or from an asymptotic
series can approximate the zeros of the parabolic cylinder function. This
problem is interesting because, like the DS equations, the polynomial sequences
have infinitely many roots while the function being approximated only has a
finite number of roots. In Sec.~\ref{s3} we show how to derive the DS equations
for a general quantum field theory. From the lowest-order calculations of the
Hermitian $\phi^4$ and non-Hermitian $-\phi^4$ theories in $D=1$, we quantify
the errors that arise and motivate the need for examining higher-order
truncations of the DS equations.

We then study the Hermitian $\phi^4$ and $\phi^6$ and the non-Hermitian $i
\phi^3$, $-\phi^4$, and $i\phi^5$ quantum field theories in $D=0$ dimensions. We
begin with the Hermitian quartic theory $\phi^4$ in Sec.~\ref{s4} and progress
to the non-Hermitian cubic theory $i\phi^3$ in Sec.~\ref{s5}, the non-Hermitian
quartic theory $-\phi^4$ in Sec.~\ref{s6}, a quintic theory in Sec.~\ref{s7},
and a sextic theory $\phi^6$ in Sec.~\ref{s8}. Conclusions are presented in
Sec.~\ref{s9}.
 
\section{Example: Zeros of a parabolic cylinder function}\label{s2}
\begin{figure}[t]
\centering
\includegraphics[scale=0.55]{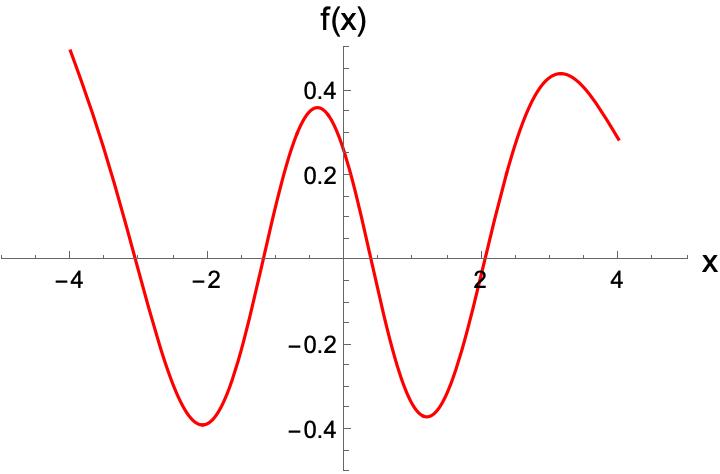}
\caption{Plot of the parabolic cylinder function $f(x)={\rm D}_{3.5}(x)$ for
$-4<x<4$; $f(x)$ is not a harmonic-oscillator eigenfunction because it blows up
when $x$ is large and negative. Note that $f(x)$ has four real zeros.}
\label{F1}
\end{figure}

To illustrate the nature of polynomial approximations, we attempt to calculate
the zeros of the parabolic cylinder function ${\rm D}_{3.5}(x)$. This function
satisfies the time-independent Schr\"odinger equation for the quantum harmonic
oscillator,
\begin{equation}
-f''(x)+\big(\fourth x^2-4\big)f(x)=0,
\label{e2.1}
\end{equation}
and is uniquely determined by the initial conditions
$${\rm D}_{3.5}(0)=\pi^{\tfrac{1}{2}}2^{\tfrac{7}{4}}/\Gamma\big(-\tfrac{5}{4}
\big)~{\rm and}~{\rm D}_{3.5}'(0)=\pi^{\tfrac{1}{2}}2^{\tfrac{9}{4}}/
\Gamma(-\tfrac{7}{4}).$$
Note that $f(x)= {\rm D}_{3.5}(x)$ is {\it not} an eigenfunction (and $4$ is not
an eigenvalue) because, as Fig.~\ref{F1} shows, while $f(x)$ vanishes as $x\to
\infty$, $f(x)$ blows up as $x\to-\infty$.

The four real zeros of $f(x)$, as shown in Fig.~\ref{F1}, are located at
\begin{equation}
-3.04735...,~~-1.19090...,~~0.39183...,~~2.04542...\,.
\label{e2.2}
\end{equation}
There are no other zeros in the complex-$x$ plane.

One way to find these zeros in (\ref{e2.2}) is to (i) expand $f(x)$ in a Taylor
series, (ii) truncate this series to obtain a polynomial, and (iii) find the
roots of the polynomial. The $2N$-term Taylor series for ${\rm D}_{3.5}(x)$ has
the form
\begin{eqnarray}
{\rm D}_{3.5}(x)&=&{\rm D}_{3.5}(0)\sum_{n=0}^{N-1}x^{2n}\frac{a_n}{(2n)!}
\nonumber\\
&&\quad+{\rm D}_{3.5}'(0)\sum_{n=0}^{N-1} x^{2n+1}\frac{b_n}{(2n+1)!}.
\label{e2.3}
\end{eqnarray}
For even powers of $x$, $a_0=1$, $a_1=-4$, and $a_n=-4a_{n-1}+\half(n-1)(2n-3)
a_{n-2}$; for odd powers of $x$, $b_0=1$, $b_1=-4$, and $b_n=-4b_{n-1}+
\half(n-1)(2n-1)b_{n-2}$.

\begin{figure}[t]
\center
\includegraphics[scale = 0.53]{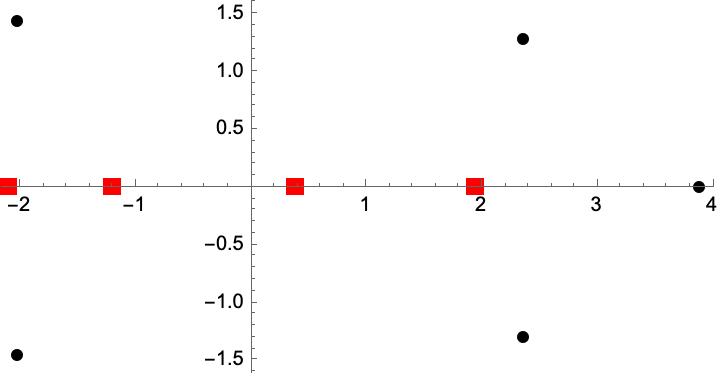}
\caption{[color online] Roots of the 9th-degree Taylor-polynomial approximation
to ${\rm D}_{3.5}(x)$ plotted in the complex-$x$ plane. Four real roots (red
squares) are located at $x=-2.09226...$, $-1.19286...$, $0.39183...$,
$1.94724...$, which are moderately close to the exact roots of $f(x)$ given in
(\ref{e2.2}). The remaining roots (black dots) are spurious zeros that gradually
move outward to complex $\infty$ as the degree of the Taylor polynomial
increases.}
\label{F2}
\end{figure}

\begin{figure}[t]
\centering
\includegraphics[scale = 0.53]{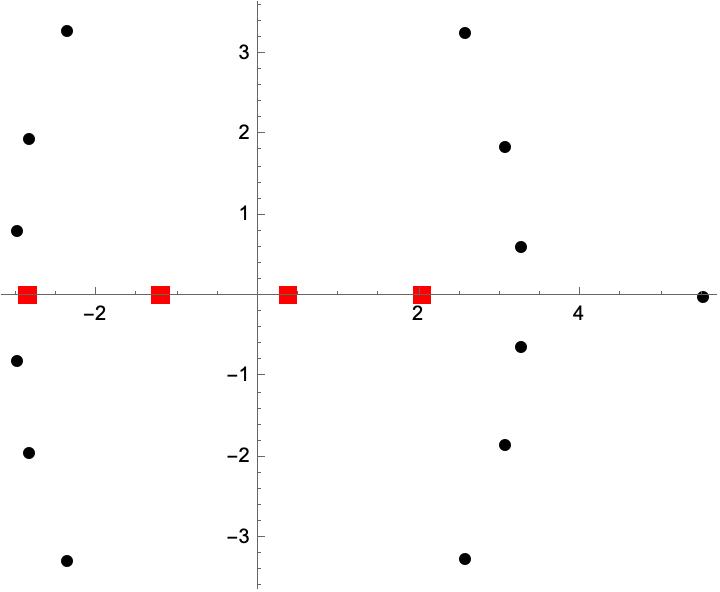}
\caption{[color online] Roots of the 17th-degree Taylor polynomial approximation
to ${\rm D}_{3.5}(x)$ plotted in the complex-$x$ plane. The real roots (red
squares) lie at $x=-2.84103...$, $-1.19090...$, $0.39183...$, $2.04519...$ and
are fairly close to their exact values in (\ref{e2.2}). Spurious roots (black
dots) lie along parenthesis-shaped curves along with an isolated spurious root
on the positive-real axis.}
\label{F3}
\end{figure}

\begin{figure}[h!]
\centering
\includegraphics[scale = 0.53]{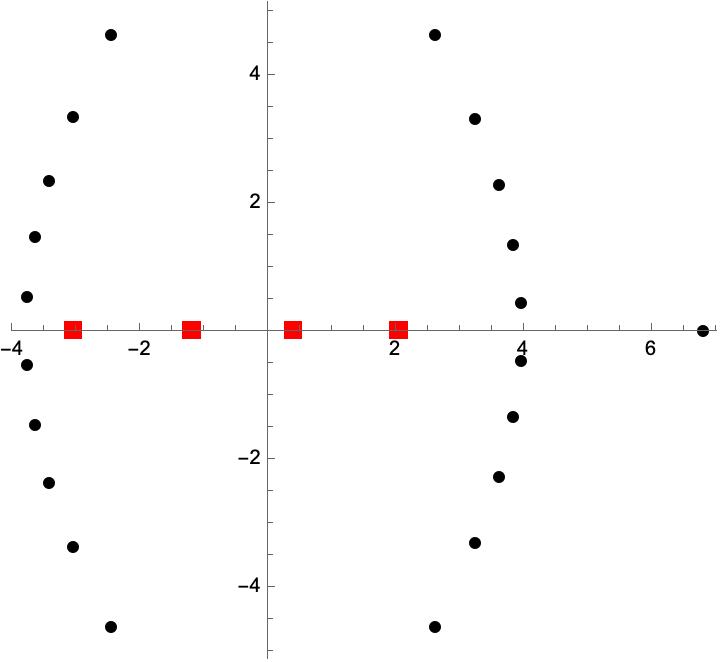}
\caption{[color online] Roots of the 25th-degree Taylor polynomial approximation
to ${\rm D}_{3.5}(x)$ plotted in the complex-$x$ plane. Real roots (red squares)
at $x=-3.04510...$, $-1.19090...$, $0.39183...$, $2.04545...$ are closer to
their exact values in (\ref{e2.2}). All but one of the spurious roots (black
dots) lie on parenthesis-shaped curves that expand outward slowly as the degree
of the Taylor polynomial increases. The isolated spurious root on the
positive-real axis also moves outward.}
\label{F4}
\end{figure}

\begin{figure}[h!]
\centering
\includegraphics[scale = 0.53]{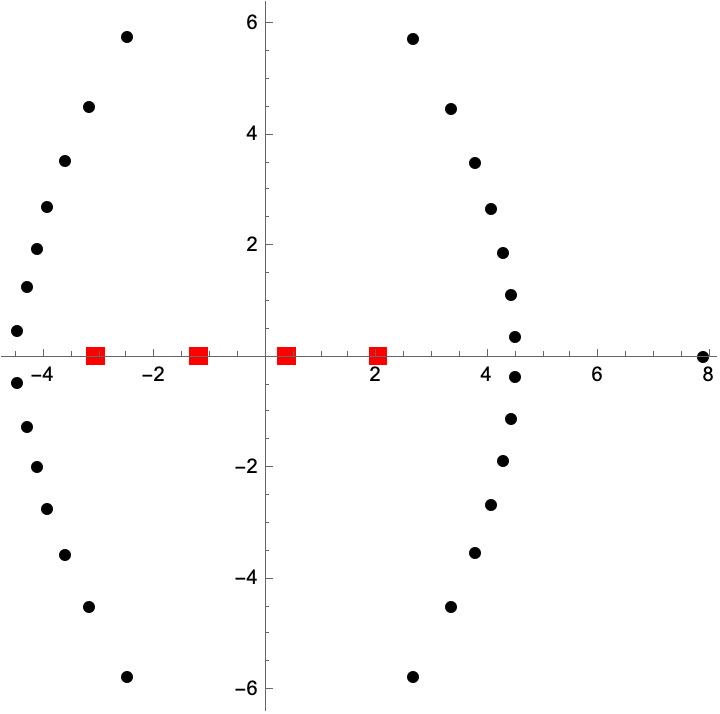}
\caption{[color online] Roots of the 33rd-degree Taylor polynomial approximation
to ${\rm D}_{3.5}(x)$ plotted in the complex-$x$ plane. Real roots (red squares)
at $x=-3.04735...$, $-1.19090...$, $0.39183...$, $2.04542...$ are now are quite
close to their exact values in (\ref{e2.2}). Spurious roots (black dots) lie on
parenthesis-shaped curves and the isolated spurious root on the positive-real
axis continue to move slowly outward.}
\label{F5}
\end{figure}

The Taylor series (\ref{e2.3}) has an infinite radius of convergence but many
terms are required to obtain accurate approximations to the zeros of ${\rm D}_{
3.5}(x)$. In Fig.~\ref{F2} we plot the zeros of the 9th-degree Taylor
polynomial. A 17th-degree Taylor polynomial gives slightly better approximations
to the zeros, as we see in Fig.~\ref{F3}. Figures \ref{F4} and \ref{F5} display
the roots of 25th-degree and 33rd-degree Taylor polynomials. As expected, the
real zeros continue to approach the exact zeros of the parabolic cylinder
function and the spurious zeros continue to move slowly outward as the degree of
the Taylor polynomial increases.

Why is such a high-degree Taylor polynomial required to provide accurate
approximations to the four zeros of the parabolic cylinder function? The answer
is that, as shown in Fig.~\ref{F1}, the parabolic cylinder function behaves
differently on the positive-real and the negative-real axes; it decays
exponentially like $\exp\big(\!-\!\fourth x^2{\big)}$ on the positive-real axis
but grows exponentially like $\exp\big(\fourth x^2\big)$ on the negative-real
axis. However, as the Taylor series converges everywhere in the complex plane,
it is difficult for the Taylor polynomials to provide accurate approximations
on both the positive and the negative axes.

Asymptotic series do not suffer from this problem because such series are not
valid in all directions in the complex plane. Their validity is limited to
wedge-shaped regions called {\it Stokes sectors}. The asymptotic series
representation for ${\rm D}_{3.5}(x)$ is
\begin{eqnarray}
{\rm D}_{3.5}(x)&\sim& e^{-x^2/4}\,x^{-7/2}\sum_{n=0}^\infty x^{-2n}\frac{c_n}
{2^n n!} \nonumber\\
&&\quad\big(|x|\to\infty,~-\threefourth\pi<{\rm arg}\,x<\threefourth\pi
\big),
\label{e2.4}
\end{eqnarray}
where $c_n=(-1)^n\Gamma\big(\frac{9}{2}\big)\Gamma\big(2n-\frac{9}{2}\big)/\pi$.
This asymptotic series is valid in a Stokes sector of angular opening
$270^\circ$ that includes the positive-real axis but not the negative-real axis.
Thus, if we factor off the leading asymptotic behavior to obtain a polynomial,
this polynomial will not give useful information about the negative zeros.

\begin{figure}[t]
\centering
\includegraphics[scale = 0.59]{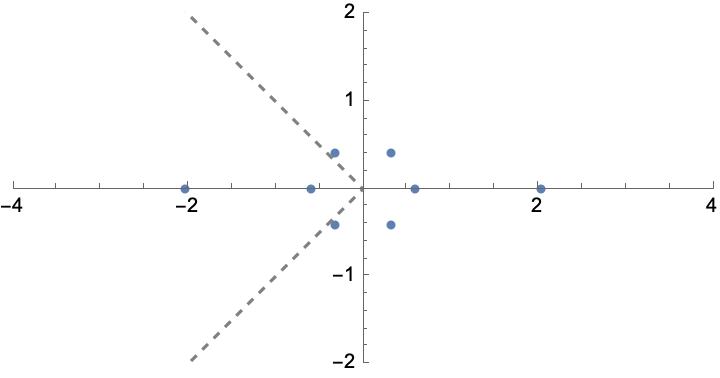}
\caption{All eight roots of the five-term polynomial obtained from the truncated
asymptotic expansion (\ref{e2.4}) plotted in the complex-$x$ plane. The two
roots on the positive axis are numerically quite accurate. The other six roots
are spurious. Dashed lines indicate the edges of the Stokes sector in which the
asymptotic series is valid.}
\label{F6}
\end{figure}

Although the asymptotic series is valid as $|x|\to\infty$, early terms provide
good approximations to the positive zeros. The positive-real roots ($x=0.59521$
and $x=2.04530$) of the five-term polynomial ($1+\alpha x^2+\beta x^4+\gamma x^6
+\delta x^8$) are already quite accurate (see Fig.~\ref{F6}); the second root is
accurate to one part in 20,000.

\begin{figure}[t]
\centering
\includegraphics[scale = 0.59]{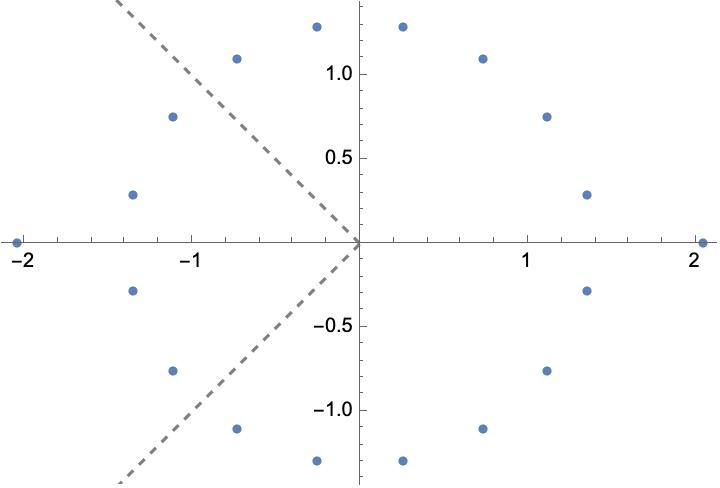}
\caption{All eighteen roots of the ten-term polynomial obtained from the
truncated asymptotic expansion (\ref{e2.4}). The roots are shown as dots in the
complex-$x$ plane. Compared with Fig.~\ref{F6}, the ring of spurious roots has
moved outward past the smaller positive zero of the parabolic cylinder function
${\rm D}_{3.5}(x)$ at $x=0.39183$ but the second positive zero is given
accurately.}
\label{F7}
\end{figure}

\begin{figure}[t]
\centering
\includegraphics[scale = 0.44]{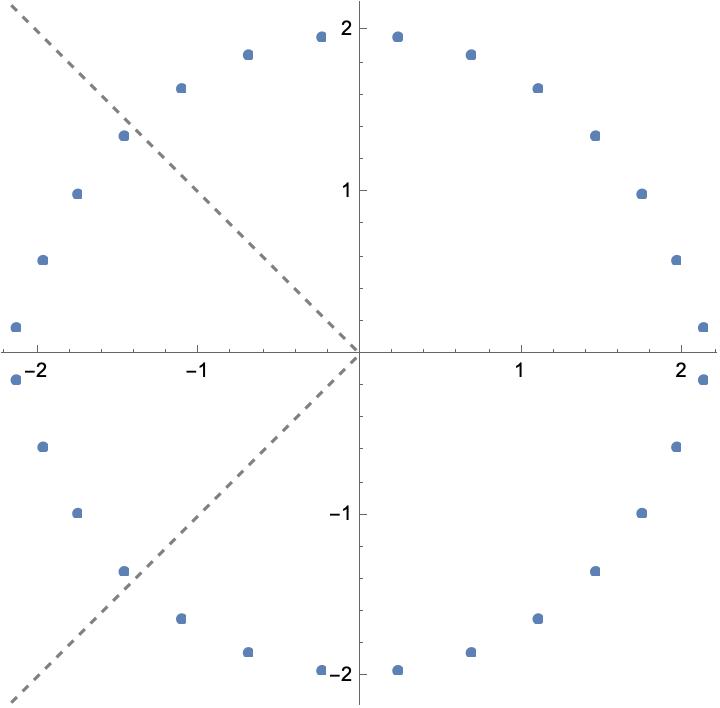}
\caption{All 28 roots of the fifteen-term polynomial obtained from the
truncated asymptotic expansion (\ref{e2.4}). The ring of spurious roots in
the complex-$x$ plane has now expanded past the actual zeros of the parabolic
cylinder function. Thus, without summation techniques the asymptotic series in
(\ref{e2.4}) is no longer useful.}
\label{F8}
\end{figure}

As the degree of the polynomial obtained from the asymptotic series increases,
the ring of spurious zeros expands. For the ten-term polynomial this ring
expands past the smaller of the two positive zeros, but there is still a very
good approximation to the larger positive root (see Fig.~\ref{F7}). For the
fifteen-term polynomial, this ring expands past the second positive zero and is
no longer directly useful (see Fig.~\ref{F8}). Summation techniques such as
Pad\'e approximation give even better accuracy but we do not discuss this here.

\begin{figure}[t]
\centering
\includegraphics[scale = 0.53]{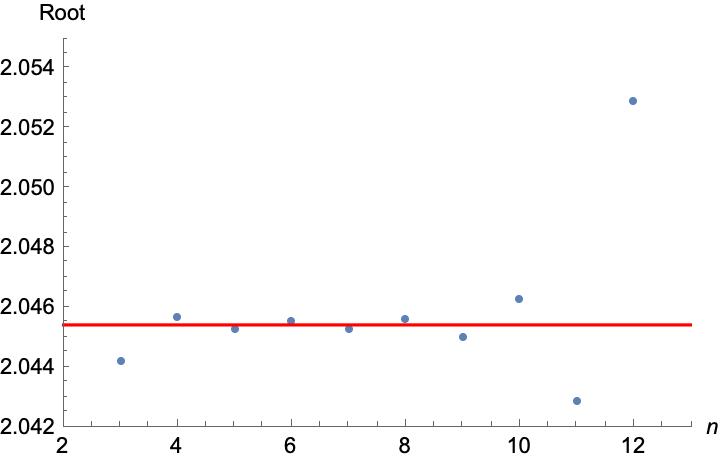}
\caption{The root near $x=2$ of the polynomial obtained from the truncated
asymptotic series approximation to the parabolic cylinder function plotted as a
function of the degree of the polynomial. The six-term polynomial gives optimal
accuracy after which the accuracy rapidly decreases.}
\label{F9}
\end{figure}

Without using summation techniques, the accuracy of an asymptotic-series
approximation typically increases as we include more terms until it reaches an
optimal level and then it decreases. This is illustrated in Fig.~\ref{F9}, which
shows the value of the root of the asymptotic-series polynomial near 2 as a
function of the number of terms in the polynomial. Note that the root oscillates
about the exact zero of the parabolic cylinder function. Optimal accuracy is
attained for the 6-term polynomial after which the accuracy decreases rapidly.

To summarize these findings, if we use a Taylor expansion to determine the roots
of the parabolic cylinder function, we find more roots than the function
actually has, and their number increases with the order of the expansion. Most
of these spurious roots are imaginary but there is at least one real spurious
root. To distinguish between actual and spurious roots one can use the criterion
of stability; that is, one can argue that the spurious roots move outward in the
complex plane while the positions of the actual roots stabilize as the order of
the expansion increases. Finally, the order of approximation that is required to
obtain an accurate result is high, which is unfortunate. We emphasize that the
coefficients of the Taylor expansion remain unchanged as we go to higher order.
This is {\it not} the case for the polynomials associated with the DS equations.

The asymptotic series approach also has advantages and disadvantages. Its region
of validity is limited to the interior of a Stokes sector and not the entire
complex plane. Thus, the number of roots that it can possibly find is also
limited. However, in its region of validity, the convergence is fast and
requires only very few terms. Like the Taylor expansion, the asymptotic series
also has many other roots in the complex plane that are spurious.

Evidently, without prior knowledge of a function, it may be difficult to
determine from a polynomial {\it expansion} of that function which roots are
close to the actual roots and which roots are spurious. In the following
sections, we restrict our analysis to quantum field theories in $D=0$ because we
can find analytic solutions. This allows us to investigate the systematics of
finding the correct roots from the polynomial DS equations.

\section{Derivation of DS Equations}\label{s3}
The objective in quantum field theory is to calculate the Green's functions
$\gamma_n(x_1,\,x_2,\,...,\,x_n)$, which are defined as vacuum-expectation
values of time-ordered products of the field $\phi(x)$:
$$\gamma_n(x_1,x_2,... x_n)\equiv\langle 0|{\rm T}\big\{\phi\big(x_1\big)
\phi\big(x_2\big)...\phi\big(x_n\big)\big\}|0\rangle.$$
These Green's functions are then combined into structures called {\it cumulants}
that give the {\it connected} Green's functions $G_n(x_1,x_2,... x_n)$. The
connected Green's functions are correlation functions that contain the physical
content (energy spectrum, scattering amplitudes) of the quantum field theory. In
principle, the program is first to solve the field equations (which are partial
differential equations like the classical equations of fluid mechanics) for the
quantum field $\phi(x)$ and then to calculate the vacuum expectation values of
products of the fields directly.

It is advantageous to calculate the connected Green's functions $G_n$, rather
than the nonconnected Green's functions $\gamma_n$ because this eliminates the
problem of vacuum divergences. As a consequence of translation invariance, each
disconnected contribution to $\gamma_n$ introduces an additional factor of the
spacetime volume $V$, which is an infinite quantity when $D>0$.

The difficulty in quantum field theory is that the field $\phi(x)$ is an {\it
operator-valued distribution} rather than a function. Free fields obey linear
differential equations but interacting fields obey {\it nonlinear} differential
equations. (The field equation for a $g\phi^4$ quantum field theory contains a
cubic term.) Unfortunately, products of fields are singular and require great
care to define them properly.

An early approach to this difficulty was to calculate Green's functions in terms
of Feynman diagrams. This perturbative procedure (in powers of the coupling
constant $g$) avoids high-level mathematical analysis and reduces the problem to
the evaluation of integrals. Indeed, in the early days of quantum field theory
one view was that one could simply {\it define} a field theory as nothing but a
set of Feynman rules and thereby avoid technical mathematical problems
\cite{R3}.

However, Feynman perturbation theory has its own mathematical difficulties:
First, individual terms in the graphical expansion may be infinite and must be
renormalized to remove the infinities. Second, the resulting renormalized
perturbation series is divergent and may not be easily summable. Third,
nonperturbative effects are difficult or even impossible to obtain by using
perturbative graphical methods alone.

Dyson and Schwinger developed another technique for calculating the Green's
functions that requires only c-number functional analysis (differential and
integral equations), so one need not be concerned about operators, Hilbert
spaces, and other mathematical issues \cite{R4,R5,R6}. In principle, one can use
this technique to obtain the nonperturbative as well as the perturbative
behavior of Green's functions. The procedure is to (i) construct an infinite
system of coupled equations called {\it Dyson-Schwinger} ({\it DS}) {\it
equations} that is satisfied exactly by the connected Green's functions, and
then (ii) truncate the infinite set of equations to a finite closed system of
coupled equations that can be solved to provide approximations to the first few
connected Green's functions.

To be precise, the DS equations are an infinite triangular system of coupled
equations obeyed by the connected Green's functions $G_n$. Each new equation
introduces additional Green's functions so a truncation of the system always
contains more Green's functions than equations and the truncated system is
underdetermined. An unbiased solution strategy is to close the truncated system
by setting the highest Green's function (or Green's functions) to zero. The
system can then be solved by successive elimination. The question investigated
here is whether this procedure gives increasingly accurate approximations to the
Green's functions as the size of the truncated system increases. We also examine
the differences between Hermitian and non-Hermitian theories. We will see below
that the accuracy of a first-order calculation of $G_2$ is significantly higher
for a one-dimensional Hermitian $\phi^4$ theory than for a non-Hermitian
$\cPT$-symmetric $-\phi^4$ theory.

The DS equations for a quantum field theory can be derived directly from the
Euclidean functional integral
\begin{equation}
Z[J]=\int\!\cD\phi\,\exp\bigg[\int\! dx\{-\cL[\phi(x)]+J(x)\phi(x)\}\bigg],
\label{e3.1}
\end{equation}
where $\cL$ is the Lagrangian and $J$ is a c-number external source. Here,
$Z[0]$ is the Euclidean partition function and $\langle0_+|0_-\rangle_J\equiv
Z[J]$ represents the {\it vacuum-persistence amplitude}; that is, the
probability amplitude for the ground state in the far past to remain in the
ground state in the far future despite the action of the external source $J$.

The vacuum-persistence functional is a generating function for the Green's
functions. If we take $n$ functional derivatives of $Z[J]$ with respect to $J$
and then set $J\equiv0$, we obtain the $n$-point Green's function $\gamma_n$:
$$\gamma_n(x_1,\,...\,x_n)=\frac{\delta}{\delta J(x_1)}...\frac{\delta}{\delta
J(x_n)}Z[J]\Big\vert_{J\equiv0}.$$
And, if we take $n$ functional derivatives of $\log(Z[J])$ with respect to $J$
and set $J\equiv0$, we obtain the {\it connected} $n$-point Green's function:
\begin{equation}
G_n(x_1,\,...\,x_n)=\frac{\delta}{\delta J(x_1)} ...\frac{\delta}{\delta J(x_n)}
\log(Z[J])\big|_{J\equiv0}.
\label{e3.2}
\end{equation}

\subsection{Example: Hermitian quartic theory in $D=1$}
For a Hermitian massless $\phi^4$ theory in one-dimensional spacetime, we begin
with the Euclidean functional integral $Z[J]=\int D\phi e^{-\int dt\cL}$, where
\begin{equation}
\cL=\half{\dot\phi}^2+\fourth g\phi^4-J\phi\quad(g>0).
\label{e3.3a}
\end{equation}
The field equation for this theory is $-{\ddot\phi}(t)+g\phi^3(t)-J(t)=0$. We
take the vacuum expectation value of the field equation and divide by $Z[J]$:
\begin{equation}
-{\ddot G}_1(t)+g\gamma_3(t,t,t)/Z[J]=J(t),
\label{e3.3}
\end{equation}
where $G_1(t)$ and $\gamma_3(t,t,t)$ are functionals of $J$.

To obtain the DS equations for the connected Green's functions we eliminate the
nonconnected Green's function $\gamma_3$ in (\ref{e3.3}) in favor of connected
Green's functions. We functionally differentiate the equation $\gamma_1(t)=
\langle 0|\phi(t)|0\rangle=Z[J]G_1(t)$ repeatedly with respect to $J(t)$:
$$\gamma_2(t,t)=\langle 0|\phi^2(t)|0\rangle=Z[J]G_2(t,t)+Z[J]G_1^2(t),$$
\begin{eqnarray}
\gamma_3(t,t,t)&=&\langle 0|\phi^3(t)|0\rangle\nonumber \\
&=&Z[J]G_3(t,t,t)+3Z[J]G_1(t)G_2(t,t)\nonumber \\
&&\quad+Z[J]G_1^3(t).\nonumber
\end{eqnarray}
We then divide this equation by $Z[J]$ and use the result to eliminate
$\gamma_3$ in (\ref{e3.3}):
\begin{eqnarray}
&&-{\ddot G}_1(t)+g[G_3(t,t,t)+3G_1(t)G_2(t,t)\nonumber\\
&&\qquad\qquad +G_1^3(t)]=J(t).
\label{e3.4}
\end{eqnarray}

This is the key equation; the entire set of DS equations is obtained from
(\ref{e3.4}) by repeated differentiation with respect to $J$ and setting
$J\equiv0$. To get the first DS equation we set $J\equiv0$ in (\ref{e3.4}). This
restores translation invariance, so $G_1$ is a constant and ${\ddot G}_1=0$.
Parity invariance implies that all odd-numbered Green's functions vanish. Thus,
the first DS equation becomes trivial: $0=0$.

To get the second DS equation we functionally differentiate (\ref{e3.4}) once
with respect to $J(s)$, set $J\equiv0$, and drop all odd-numbered Green's
functions:
\begin{equation}
-{\ddot G}_2(s-t)+M^2 G_2(s-t)+gG_4(s,t,t,t)=\delta(s-t),
\label{e3.5}
\end{equation}
where the renormalized mass is
\begin{equation}
M^2=3gG_2(0).
\label{e3.6}
\end{equation}

We cannot solve (\ref{e3.5}) because it is one equation in two unknowns, $G_2$
and $G_4$. As stated above, each new DS equation introduces one new unknown
Green's function: The third DS equation is trivial but the fourth contains
$G_6$, the fifth is trivial but the sixth contains $G_8$, and so on. To proceed,
we simply set $G_4=0$ in (\ref{e3.5}).

To solve the resulting equation we take a Fourier transform to get $(p^2+M^2)
{\tilde G}_2(p)=1$. Thus, the two-point connected Green's function in momentum
space is
$${\tilde G}_2(p)=1/\big(p^2+M^2\big).$$
Taking the inverse transform, we get $G_2(t)=e^{-M|t|}/(2M)$, so $G_2(0)=1/
(2M)$. Inserting $G_2(0)$ into (\ref{e3.6}) gives a cubic equation for the
renormalized mass whose solution for $g=1$ is $M=(3/2)^{1/3}=1.145...\,$.

To check the accuracy of this result we note that the renormalized mass is the
energy of the lowest excitation above the ground state. For this model (massless
quantum anharmonic oscillator) the exact answer is $M=E_1-E_0=1.088...\,$. Thus,
the DS result is 5.2\% high, which is not bad for a leading-order truncation.

\subsection{Example: $\cPT$-symmetric quartic theory in $D=1$}
We obtain a non-Hermitian $\cPT$-symmetric massless $\phi^4$ theory in $D=1$
if $g$ in (\ref{e3.3a}) is negative. In this case the Green's functions are not
parity symmetric, so the odd-$n$ Green's functions do not vanish. The first DS
equation is not trivial, $3G_2(0)+G_1^2=0$, where we have divided by the common
factor $G_1$. Following the procedure in the example above, the second DS
equation leads to two more equations
$$M^2=3g[G_1^2+G_2(0)],\quad G_2(0)=1/(2M).$$
We set $g=-1$ and solve the three equations above for the renormalized mass:
\begin{equation}
M=3^{1/3}=1.442...\,.
\label{e3.8}
\end{equation}
The exact value of $M$ obtained by solving the Schr\"odinger equation for the
$\cPT$-symmetric quantum-mechanical Hamiltonian $H=\half p^2-\fourth x^4$ is
$M=E_1-E_0=1.796...\,$. Thus, the result in (\ref{e3.8}) is 19.7\% low.

The two examples above raise the following question: Does the accuracy improve
if we perform higher-level truncations of the DS equations? In general, this is
not an easy question to answer because higher-order truncations of the DS
equations lead to nonlinear integral equations, which require detailed numerical
analysis. However, we can solve the DS equations in very high order to study the
convergence in zero spacetime dimensions. In the next sections we examine this
question in detail for the $D=0$ Hermitian $g\phi^4$ ($g>0$) theory, the $D=0$
non-Hermitian $i\phi^3$ theory, the $D=0$ non-Hermitian $g\phi^4$ ($g<0$)
theory, the $D=0$ non-Hermitian $-i\phi^5$ theory, and the $D=0$ Hermitian
$\phi^6$ theory.

\section{$D=0$ Hermitian quartic theory}\label{s4}
In zero-dimensional spacetime the functional integral (\ref{e3.1}) becomes the
ordinary integral
\begin{equation}
Z[J]=\int_{-\infty}^\infty d\phi\,e^{-\cL(\phi)},
\label{e4.1}
\end{equation}
where $\cL(\phi)=\fourth\phi^4-J\phi$ and we have set $g=1$. The connected
two-point Green's function is an ordinary integral, which we evaluate exactly:
\begin{eqnarray}
G_2&=&\int_{-\infty}^\infty d\phi\,\phi^2 e^{-\phi^4/4} \Big/
\int_{-\infty}^\infty d\phi\,e^{-\phi^4/4} \nonumber \\
&=&2\Gamma\big(\threefourth\big)\big/\Gamma\big(\fourth\big)
= 0.675\,978...\,.
\label{e4.2}
\end{eqnarray}

The theory defined in (\ref{e4.1}) has parity invariance when $J=0$, so all odd
Green's functions vanish, $G_1=G_3=G_5=\,...\,=0$ and the first nontrivial DS
equation is $G_4=-3G_2^2+1$. If we truncate this equation by setting $G_4=0$ and
solve the resulting equation $3G_2^2=1$, we get the approximate numerical result
$G_2=1/\sqrt{3}=0.577\,350...\,$. In comparison with (\ref{e4.2}) this result is
14.6\% low.

Let us include more DS equations: The first four are
\begin{eqnarray}
G_4\! &=& \!-3G_2^2+1,\nonumber\\
G_6\! &=& \!-12G_2G_4 - 6 G_2^3,\nonumber\\
G_8\! &=& \!-18G_2G_6-30G_4^2-60G_2^2G_4,\nonumber\\
G_{10}\! &=& \!-24G_2G_8-168G_4G_6-126G_2^2G_6-420G_2G_4^2,\nonumber
\end{eqnarray}
and the next six are
\begin{eqnarray}
G_{12}\! &=& \!-30G_2G_{10}-360G_4G_8-216G_2^2G_8-378G_6^2 \nonumber\\
&&\!-3024G_2G_4G_6,\nonumber\\
G_{14}\! &=& \!-36G_2G_{12}-660G_4G_{10}-330G_2^2G_{10}\nonumber\\
\!&&\!  -2376G_6G_8-7920G_2G_4G_8-8316G_2G_6^2\nonumber\\
&&\! -41580G_3G_5G_6-27720G_4^2G_6,\nonumber\\
G_{16}\! &=& \!-42G_2G_{14}-1092G_4G_{12}-468G_2^2G_{12}\nonumber\\
&&\!-6006G_6G_{10}-17160G_2G_4G_{10}-5148G_8^2\nonumber\\
&&\!-61776G_2G_6G_8-102960G_4^2G_8-216216G_4G_6^2,\nonumber\\
G_{18}\! &=&\!-48G_2G_{16}-1680G_4G_{14}-630G_2^2G_{14}\nonumber\\
&&\! -13104G_6G_{12}-32760G_2G_4G_{12}-34320G_8G_{10}\nonumber\\
&&\! -180180G_2G_6G_{10}-300300G_4^2G_{10}\nonumber\\
&&\! -154440G_2G_8^2 -2162160G_4G_6G_8-756756G_6^3,\nonumber\\
G_{20}\! &=& \!-54G_2G_{18}-2448G_4G_{16}-816G_2^2G_{16}\nonumber \\
&& \! -25704G_6G_{14}-57120G_2G_4G_{14}-95472G_8G_{12}\nonumber\\
&&\! -445536G_2G_6G_{12}-742560G_4^2G_{12}-72930G_{10}^2\nonumber\\
&&\! -7001280G_4G_8^2-1166880G_2G_8G_{10}\nonumber\\
&&\! -8168160G_4G_6G_{10}-14702688G_6^2G_8\nonumber\\
&&\! -17153136G_6G_7^2,\nonumber\\
G_{22}\! &=& \!-60G_2G_{20}-3420G_4G_{18}-1026G_2^2G_{18}\nonumber\\
&&\! -46512G_6G_{16}-93024G_2G_4G_{16}\nonumber\\
&&\! -232560G_8G_{14}-976752G_2G_6G_{14}\nonumber\\
&&\! -1627920G_4^2G_{14}-503880G_{10}G_{12}\nonumber\\
&&\! -3627936G_2G_8G_{12} -25395552G_4G_6G_{12}\nonumber\\
&&\! -2771340G_2G_{10}^2-66512160G_4G_8G_{10}\nonumber\\
&&\! -69837768G_6^2G_{10}-119721888G_6G_8^2.
\label{e4.4}
\end{eqnarray}

Because the DS equations (\ref{e4.4}) are {\it exact}, we can find the precise
values of {\it all} $G_{2n}$ sequentially by substituting the exact value of
$G_2$ from (\ref{e4.2}) into (\ref{e4.4}). The results are given in Table
\ref{t1}. Observe that the $G_{2n}$ alternate in sign as $n$ increases, a
feature that is not immediately evident from the structure of the equations in
(\ref{e4.4}). Close examination of the terms contributing to a given $G_{2n}$
reveals that all terms are of similar size, so it is not easy to identify a
dominant contribution. 

The oscillation in sign of $G_{2n}$ as $n$ increases differs from the behavior
of the {\it disconnected} Green's functions, 
\begin{equation}
\gamma_{2n}=\int_{-\infty}^\infty d \phi\phi^{2n}e^{-\phi^4/4}=2^{n-1/2}\Gamma
(\tfrac{2n+1}4),
\label{e4.71}
\end{equation}
all of which from (\ref{e4.71}) are positive. The first eleven numerical values
are also given in Table~\ref{t1}.

\begin{table}[h!]
\caption{Exact values of the first 11 nonzero connected Green's functions 
(left) and the first 11 nonzero disconnected Green's functions (right) for the
Hermitian quartic theory (\ref{e4.1}).}
\centering{
\begin{tabular}{|c|c|}
\hline\hline
$G_2^{\rm exact}$ & $0.675\,978\,24$\\
$G_4^{\rm exact}$ & $-0.370\,839\,74$\\
$G_6^{\rm exact}$ & $1.154\,839\,49$\\
$G_8^{\rm exact}$ & $-8.010\,060\,86$\\
$G_{10}^{\rm exact}$ & $96.364\,571\,49$\\
$G_{12}^{\rm exact}$ & $-1,775.987\,088\,64$\\
$G_{14}^{\rm exact}$ & $46,449.956\,507\,74$\\
$G_{16}^{\rm exact}$ & $-1,635,683.384\,912\,06$\\
$G_{18}^{\rm exact}$ & $74,607,360.536\,889\,26$\\
$G_{20}^{\rm exact}$ & $-4,278,841,318.741\,397\,69$\\
$G_{22}^{\rm exact}$ & $301,366,607,264.871\,591\,99$\\
\hline
\end{tabular}
\hspace{1.1cm}
\begin{tabular}{|c|c|}
\hline\hline
$\gamma_2^{\rm exact}$ & $1.733$\\
$\gamma_4^{\rm exact}$ & $2.56369$\\
$\gamma_6^{\rm exact}$ & $5.199$\\
$\gamma_8^{\rm exact}$ & $12.8185$\\
$\gamma_{10}^{\rm exact}$ & $36.393$\\
$\gamma_{12}^{\rm exact}$ & $115.366$\\
$\gamma_{14}^{\rm exact}$ & $400.323$\\
$\gamma_{16}^{\rm exact}$ & $1\,499.76$\\
$\gamma_{18}^{\rm exact}$ & $6\,004.85$\\
$\gamma_{20}^{\rm exact}$ & $25\,495.9$\\
$\gamma_{22}^{\rm exact}$ & $114\,092.$\\
\hline
\end{tabular}
}
\label{t1}
\end{table}

It is possible to check the expressions for the connected Green's functions in
(\ref{e4.4}) using an alternative, independent method. We calculate the $G_{2n}$
directly from a generating function $w(x)$, which in this case is possible,
since we know $\gamma_{2n}$ explicitly: we can write down the generating
function for $G_{2n}$ in terms of it,
\begin{equation}
w(x)=\ln\big[1+\frac{1}{2!}\frac{\gamma_2}{\gamma_0}x^2+\frac{1}{4!}\frac{
\gamma_4}{\gamma_0}x^4+\frac{1}{6!}\frac{\gamma_6}{\gamma_0}x^6+\dots\big],
\label{e4.72}
\end{equation}
expand this in a Taylor series, and identify the $G_{2n}$ as $(2n!)\times$ the
coefficient of $x^{2n}$. One easily finds that the coefficient of $x^2$ is
$\Gamma(\tfrac 34)/\Gamma(\tfrac14)$, so that one recovers the value of $G_2$
given in (\ref{e4.2}). The next five Green's functions calculated in this way
are
\begin{eqnarray}
G_4 &=& 1-12 \frac{\Gamma(\tfrac 3 4)^2}{\Gamma(\tfrac 14)^2},\nonumber\\
G_6 &=&-24\frac{\Gamma(\tfrac 3 4)}{\Gamma(\tfrac1 4)}+240
\frac{\Gamma(\tfrac 3 4)^3}{\Gamma(\tfrac1 4)^3} ,\nonumber\\
G_8 &=& -30 + 1\,344\frac { \Gamma(\tfrac 3 4)^2 }{\Gamma(\tfrac1 4)^2}- 
10\,080 \frac{\Gamma(\tfrac 3 4)^4}{\Gamma(\tfrac1 4)^4},\nonumber\\
G_{10}&=& 4\,632 \frac{\Gamma(\tfrac 3 4)}
{\Gamma(\tfrac 14)}- 120 \,960 \frac{\Gamma(\tfrac 3 4)^3}
{\Gamma(\tfrac 14)^3} + 725\,760\frac{\Gamma(\tfrac 3 4)^5}
{\Gamma(\tfrac 14)^5}.\nonumber\\
G_{12}&=& 9\,120 - 877\,536 \frac{\Gamma(\tfrac 3 4)^2}
{\Gamma(\tfrac 14)^2} + 15\,966\,720 \frac{\Gamma(\tfrac 3 4)^4}
{\Gamma(\tfrac 14)^4}\nonumber\\
&&\quad - 79\,833\,600\frac{\Gamma(\tfrac 3 4)^6}
{\Gamma(\tfrac 14)^6},
\label{e4.73}
\end{eqnarray}
with the expansion of the logarithm leading to alternating signs of the terms 
contributing to each $G_{2n}$.

A numerical evaluation of (\ref{e4.73}) confirms the exact values given in Table
\ref{t1}. The analytic relationships among the $G_{2n}$, as derived from the DS
equations (\ref{e4.4}), can be easily confirmed. This calculation confirms the
DS equations, but (\ref{e4.73}) does not lend itself to an asymptotic analysis
because the sign of $G_{2n}$ as evaluated from these expressions is determined
by a delicate cancellation of terms having different signs. The alternation in
signs of the $G_{2n}$ occurs because these functions are cumulants, reflecting
only connected terms and therefore requiring subtractions. From the DS equations
(\ref{e4.4}) it is not at all obvious that the signs are oscillating. 

\subsection{Approximate solutions}
As a rule, we do not know the exact solutions to the DS equations, and must
therefore employ approximate methods of solution. The system of DS equations
(\ref{e4.4}) is not closed. Rather it is triangular, and the number of unknowns 
is always one more than the number of equations. A standard unbiased procedure
is to define a truncation scheme in which as a first approximation 
$G_{4}=G_{6}=G_{8}= ... = 0$; the next level of approximation is reached by 
setting $G_{6}=G_{8}= G_{10}= ...=0$, the next by setting $G_{8}=G_{10}= 
G_{12}=G_{14}= ...=0$, and so on. 

To do this efficiently, we reorganize (\ref{e4.4}). We eliminate $G_4$ by
substituting the first equation into the second, we eliminate $G_6$ by
substituting the first two equations into the third, and so on. Continuing this
way, we obtain an expression for $G_{2n}$ as an $n$th degree polynomial in $G_2$
only. We denote the {\it monic} form of these polynomials (where the
highest power of $x$ is 1) as $P_n(G_2)$. The first ten such polynomials are
\begin{eqnarray}
P_2(x) &=& x^2-\tfrac{1}{3},\nonumber\\
P_3(x) &=& x^3-\tfrac{2}{5}x,\nonumber\\
P_4(x) &=& x^4-\tfrac{8}{15}x^2+\tfrac{1}{21},\nonumber\\
P_5(x) &=& x^5-\tfrac{2}{3}x^3+\tfrac{193}{1890}x\nonumber\\
P_6(x) &=& x^6-\tfrac{4}{5}x^4+\tfrac{277}{1575}x^2-\tfrac{76}{10395},
\nonumber\\
P_7(x) &=& x^7-\tfrac{14}{15}x^5+\tfrac{361}{1350}x^3-\tfrac{85}{3861}x,
\nonumber\\
P_8(x) &=& x^8-\tfrac{16}{15}x^6+\tfrac{356}{945}x^4-\tfrac{475792}{10135125}x^2
+\tfrac{1229}{1091475},\nonumber\\
P_9(x) &=& x^9-\tfrac{6}{5}x^7+\tfrac{529}{1050}x^5-\tfrac{13583}{160875}x^3
\nonumber\\
&&\quad+\tfrac{8413529}{1929727800}x,\nonumber\\
P_{10}(x) &=& x^{10}-\tfrac{4}{3}x^8+\tfrac{613}{945}x^6-\tfrac{92464}{675675}
x^4\nonumber\\
&&\quad+\tfrac{3658792}{328930875}x^2-\tfrac{32372}{186642225},\nonumber\\
P_{11}(x) &=&x^{11}-\tfrac{22}{15}x^9+\tfrac{7667}{9450}x^7
-\tfrac{190319}{921375}x^5\nonumber \\
&&\quad+\tfrac{130461193}{5638815000}x^3-\tfrac{61559809}{74996239500}x.
\label{e4.5}
\end{eqnarray}

Truncating the DS equations (\ref{e4.4}) is equivalent to finding the zeros of
these polynomials. We list the nonnegative zeros below (negative zeros are
excluded because $G_2=M^{-2}$, where $M$ is the renormalized mass). The first
seven sets of zeros are
\begin{eqnarray}
{\rm zero~of}~P_2:&&~0.577350,\nonumber\\
{\rm zeros~of}~P_3:&&~0.0,~~ 0.632456,\nonumber\\
{\rm zeros~of}~P_4:&&~0.336742,~~0.648026,\nonumber\\
{\rm zeros~of}~P_5:&&~0.0,~~0.488357,~~0.654350,\nonumber\\
{\rm zeros~of}~P_6:&&~0.232147,~~0.560220,~~0.657466,\nonumber\\
{\rm zeros~of}~P_7:&&~0.0,~~0.376821,~~0.597310,~~0.659212,\nonumber\\
{\rm zeros~of}~P_8:&&~0.176270,~~0.466447,~~0.618098,\nonumber\\
&&\quad ~~0.660287,\nonumber
\end{eqnarray}
and the next three sets of zeros are
\begin{eqnarray}
{\rm zeros~of}~P_9:&&~0.0,~~0.302770,~~0.523189,~~0.630624,\nonumber\\
&&\quad ~~0.660997,\nonumber\\
{\rm zeros~of}~P_{10}:&&~0.141830,~~0.392352,~~0.560204,\nonumber\\
&&\quad ~~0.638652,~~0.661493,\nonumber\\
{\rm zeros~of}~P_{11}:&&~0.0,~~0.251866,~~0.456057,~~0.585125,\nonumber\\
&&\quad~~0.644070,~~0.661853.
\label{e4.6}
\end{eqnarray}

\begin{figure}[t]
\centering
\includegraphics[scale = 0.28]{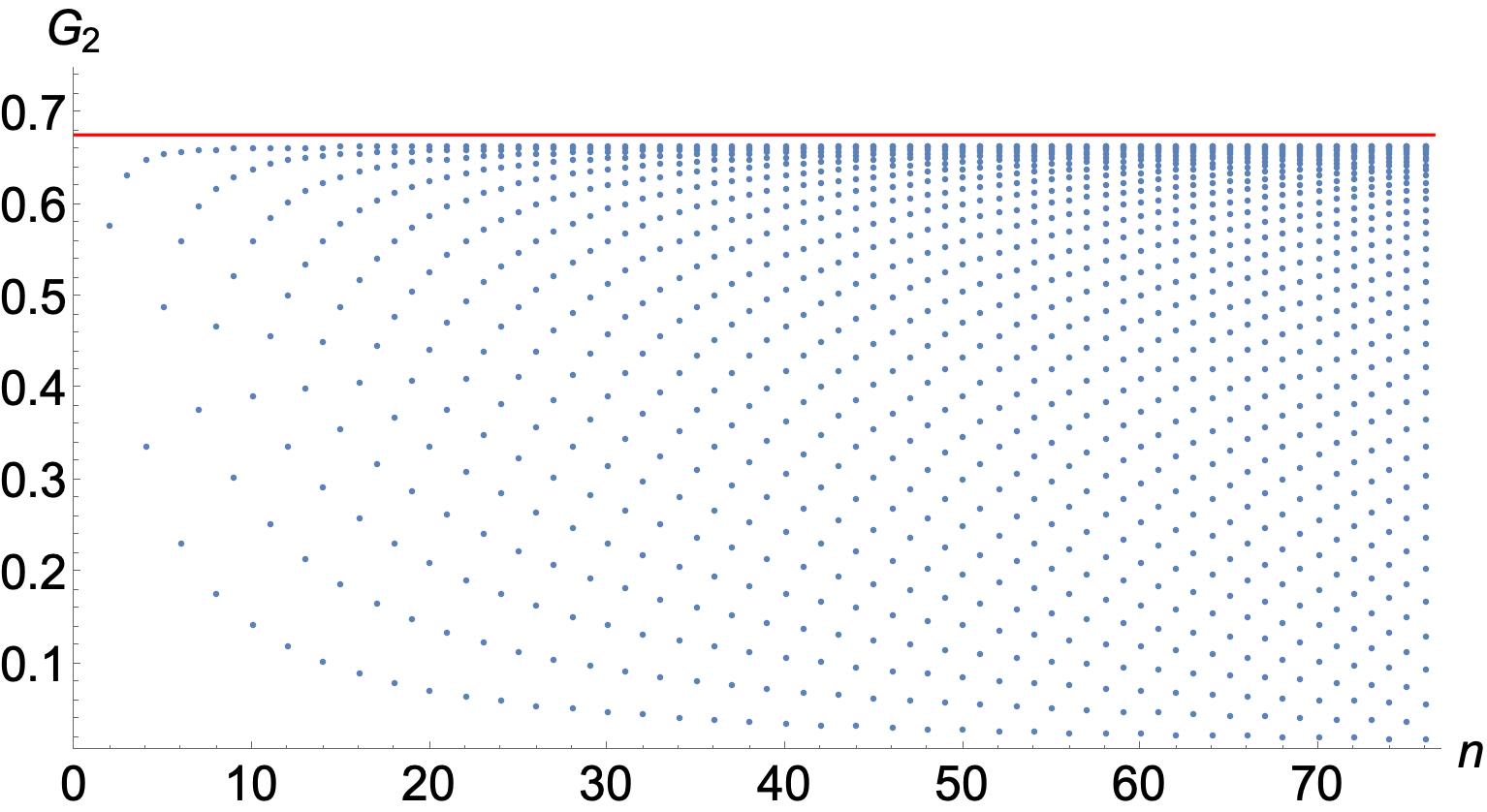}
\caption{Zeros of $P_n(x)$ in (\ref{e4.5}) plotted as a function of $n$. The
zeros are nondegenerate and range from 0 up to just below $0.675\,978...$, the
exact value of $G_2$ in (\ref{e4.2}) (heavy horizontal line) and they become
more dense at the upper end of this range. The zeros of successive polynomials
interlace.}
\label{F10}
\end{figure}

The roots up to $n=80$ are plotted in Fig.~\ref{F10}. Note that all roots are
real and nondegenerate, and range from 0 up to just below the exact value of
$G_2$ in (\ref{e4.2}). If we did not already know the exact value $G_2$, we
could not guess which root gives the best approximation to $G_2$. However, with
increasing truncation order, the roots become more dense at the upper end of the
range, so we would conjecture that the largest root gives the best
approximation. Unfortunately, while the accuracy improves monotonically with the
order of the truncation, it improves slowly; the largest root of $P(x)$ is still
1.85\% below the exact value. Using Richardson extrapolation, we can determine
the value to which the largest root converges \cite{R7}: $G_2=0.663\,488\dots$.
Thus, the limiting value of the sequence of roots does {\it not} converge to the
true value $G_2=0.675\,978\dots$ \cite{R1}. To understand this discrepancy, we
examine the large-$n$ asymptotic behavior of the $G_{2n}$ in detail in the
following subsection.

Figure \ref{F10} also shows that the zeros of successive polynomials $P_n(x)$
interlace. This interlacing behavior might suggest that the polynomials $P_n(x)$
form an orthogonal set with respect to some weight function, but this conjecture
is false. Nevertheless, these polynomials do have interesting properties. In
particular, there are relatively simple formulas for the polynomial
coefficients: The coefficient of $x^n$, the highest power of $x$ in $P_n(x)$, is
1 (these are monic polynomials) and the formula for the coefficient of
$x^{n-2}$, the second highest power of $x$, is $-\tfth n~~(n>2)$. The
coefficient of $x^{n-4}$ is
$$\textstyle{\frac{1}{2!}}(\tfth)^2[n^2-\textstyle{\frac{227}{84}}n]\quad
(n>4),$$
the coefficient of $x^{n-6}$ is
$$-\textstyle{\frac{1}{3!}}(\tfth)^3[n^3-\textstyle{\frac{227}{28}}n^2
+\textstyle{\frac{31453}{2002}}n]\quad(n>6),$$
and the coefficient of $x^{n-8}$ is
$$\textstyle{\frac{1}{4!}}(\tfth)^4[n^4-\textstyle{\frac{227}{14}}n^3
+\textstyle{\frac{28505063}{336336}}n^2-\textstyle{\frac{404875283}{2858856}}n]
\quad(n>8).$$

\subsection{Large-$n$ behavior of the Green's functions $G_{2n}$} 
The question is whether it is valid to truncate the DS equations (\ref{e4.4}) by
replacing $G_{2n}$ with zero. To answer this question we look at the asymptotic
behavior of $G_{2n}$ for large $n$. We have shown both numerically and
analytically \cite{R1} that the asymptotic behavior of $G_{2n}$ is
\begin{equation}
G_{2n} \sim 2r^{2n}(-1)^{n+1}(2n-1)!\quad(n\to\infty),
\label{e4.7}
\end{equation}
where $r=0.409\,505\,7...\,.$

To obtain this result analytically we substitute
$$G_{2n}=(-1)^{n+1}(2n-1)!\,g_{2n},$$
which is suggested by the numerical result in (\ref{e4.7}),
and we define a generating function $u(x)$ for the numbers $g_{2n}$:
\begin{equation}
u(x)\equiv x g_2+x^3 g_4+x^5 g_6 +...\,.
\label{e4.9}
\end{equation}
This generating function obeys the second-order nonlinear differential equation
$$u''(x)=3u'(x)u(x)-u^3(x)-x,$$
subject to the initial conditions $u(0)=0$ and
$$u'(0)=G_2=2\Gamma\big(\threefourth\big)/\Gamma\big
(\fourth\big)=0.675\,978\,240\,067\,285...\,.$$

The substitution $u(x)=-y'(x)/y(x)$ then gives the third-order {\it linear}
differential equation
$$y'''(x)=xy(x),$$
which is a higher-order generalization of the Airy equation $y''(x)=xy(x)$. The
function $y(x)$ satisfies the initial conditions $y(0)=1$, $y'(0)=0$, and
$$y''(0)=-G_2 =-0.675\,978\,240\,067\,285...\,.$$
The exact solution $y(x)$ satisfying these boundary conditions is found by
taking a cosine transform: 
\begin{equation}
y(x)=\frac{2\sqrt{2}}{\Gamma(1/4)}\int_0^\infty dt\,\cos(xt)\,e^{-t^4/4}.
\label{e4.11}
\end{equation}

When $y(x)$ passes through 0, $u(x)$ becomes infinite, so the value of $x$ at
which $y(x)=0$ determines the radius of convergence of the series (\ref{e4.9})
for the generating function. We find that $u(x)$ passes through 0 at $x=\pm
2.441\,968...\,$. Therefore, $r=1/x=0.409\,506...\,$, which confirms the
numerical results in (\ref{e4.7}).

The asymptotic behavior in (\ref{e4.7}) is surprising; it shows that the
connected Green's functions $G_{2n}$ grow much faster with increasing $n$ than
the nonconnected Green's functions $\gamma_{2n}$ which are given exactly for all
$n$ in (\ref{e4.71}). One might not expect $G_{2n}$ to grow faster than
$\gamma_{2n}$ because we obtain the connected Green's function by {\it
subtracting} the disconnected parts from $\gamma_{2n}$. Surprisingly,
subtracting disconnected parts makes the absolute values of the connected
Green's functions larger and not smaller with increasing $n$.

Even more remarkable is that neglecting the huge quantity $G_{2n}$ on the left
side of the truncated DS equations (\ref{e4.4}) still leads to a reasonably
accurate result for $G_2$, as Fig.~\ref{F10} shows. This accuracy {\it improves}
with increasing $n$. We can begin to understand this heuristically by observing
that while the term on the left side is very big, the terms on the right side
are of roughly comparable size because the coefficients are also big.

The numerical technique of Legendre interpolation provides a helpful analogy.
Given a set of $n$ data points $x_1,\,...,\,x_n$ at which we measure a function
$f(x)$, 
$$f(x_1)=f_1,\,...,\,f(x_n)=f_n,$$
Legendre interpolation fits this data by constructing a polynomial $P_{n-1}(x)$
of degree $n-1$ that passes exactly through the value $f(x_k)$ at $x=x_k$ for
all $1\leq k\leq n$. There is a simple formula for this polynomial. However,
this construction has a serious problem; while the constructed polynomial passes
exactly through the data points, between data points the polynomial exhibits
wild oscillations where it becomes alternately large and positive and large and
negative. This reveals a fundamental instability associated with high-degree
polynomials. This instability is associated with the inherent stiffness of
polynomials \cite{R8}. If there are many data points, it is much better to use a
{\it least-squares} polynomial approximation, which passes close to, but not
exactly through the input data points. (This explains why {\it cubic} splines
are used to approximate functions rather than, say, octic splines.)

It is precisely the instability associated with the stiffness of high-degree
polynomials that allows the DS approach to give reasonably accurate results! If
we use the exact values of the Green's functions on the right side of the DS
equations (\ref{e4.4}), we obtain the exact value of the Green's function on the
left side, which is a huge number. However, changing the Green's functions on
the right side of (\ref{e4.4}) very slightly by replacing the exact values by
the {\it approximate} values of the lower Green's functions now gives 0, instead
of $G_{2n}$.

Pad\'e approximation does not improve the calculation of $G_2$ from the DS
polynomials in (\ref{e4.5}). One might anticipate that Pad\'e techniques would
be useful because the coefficients of successive powers of $x$ alternate in
sign. The approach would be to divide all odd-numbered polynomials by $x$ and
then to replace $x^2$ in each polynomial by $y$. If one does this for $P_{11}$,
for example, one can then calculate the $[1,4]$, $[2,3]$, $[3,2]$, and $[4,1]$
approximants. Unfortunately, the zeros of these approximants are not near the
exact value of $G_2^2$, and such an attempt to improve the accuracy of the DS
equations fails. Why does this approach fail? Pad\'e approximation accelerates
the convergence of a truncated series even if the series diverges. However,
unlike the infinite Taylor series expansion of the parabolic cylinder function
in (\ref{e2.3}) where the coefficients of powers of $x$ remain the {\it same} as
the order is increased, the coefficients in the DS equations {\it change from
order to order}. Other approaches, such as assuming a value of $G_{2n}$
estimated from $G_{2n-2}$ converge to a limit that is very slightly closer to
the correct one, but which is still not correct, and thus also fail.

\begin{figure}[h!]
\centering
\includegraphics[scale = 0.23]{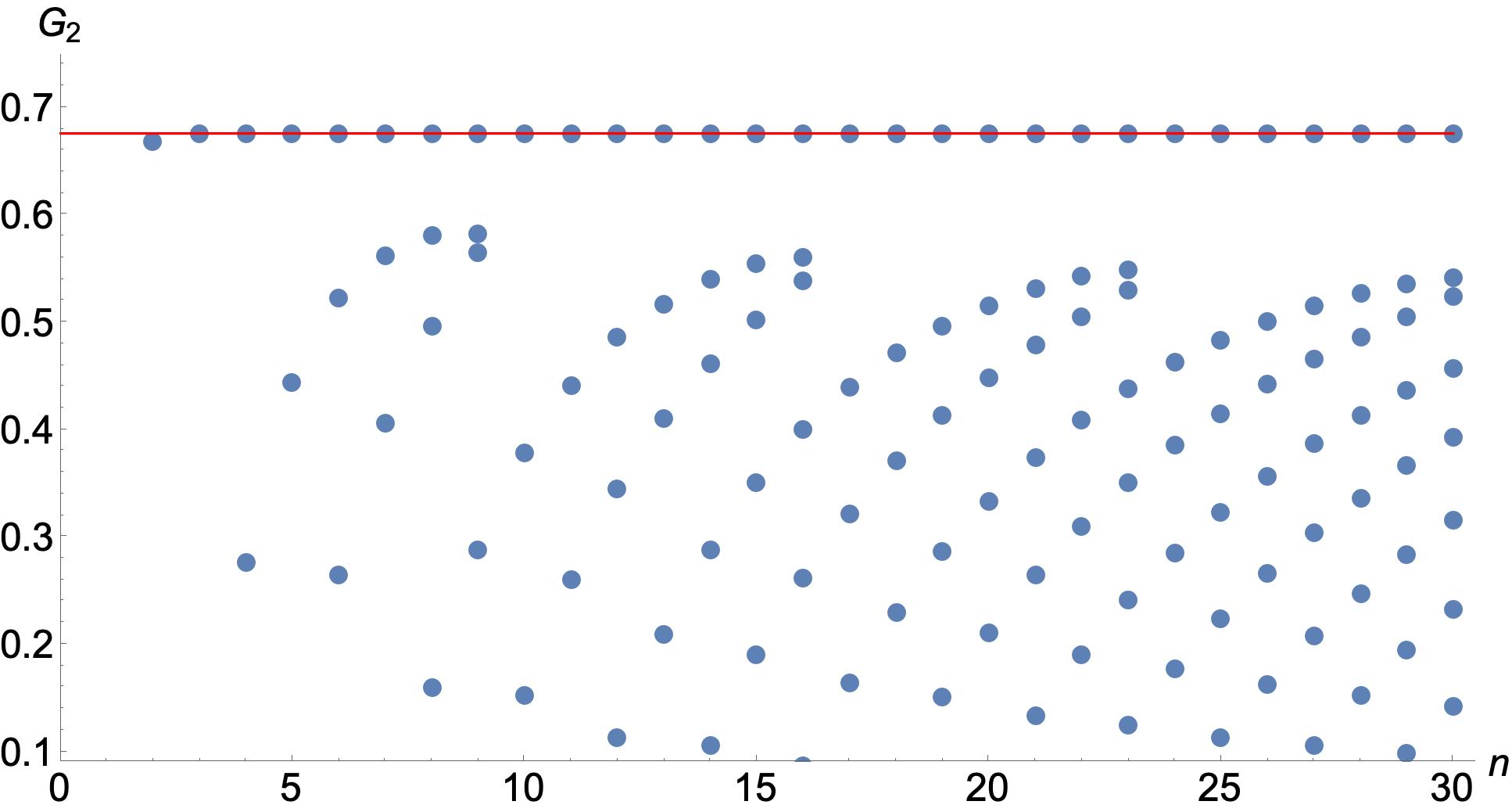}
\caption{Results for $G_2$ plotted as a function of $n$ ($1\leq n\leq 30$)
calculated using the asymptotic approximation for $G_{2n}$ in (\ref{e4.7}).
We observe a dramatic improvement over Fig.~\ref{F10}; there is no longer a
dense concentration of roots below the exact answer but rather an isolated root
that is six orders of magnitude closer to the exact answer.}
\label{F11}
\end{figure}

One approach {\it does} give excellent numerical results: If the left side of
the DS equations is approximated by the asymptotic approximation (\ref{e4.7}),
$G_2$ reaches an accuracy of seven decimal places in only six steps. (See
Fig.~\ref{F11}.) At $n=7$, we have $G_2=0.675\,978\,218\ldots$ in comparison
with the exact result $G_{2,\,{\rm exact}}=0.675\,978\,240\ldots\,.$

While we have gained six orders of magnitude in precision, the result in
Fig.~\ref{F11} is not exact. This is because (\ref{e4.7}) is only a {\it
leading-order} asymptotic approximation. Higher-order asymptotic approximations
for $G_{2n}$ will improve this impressive numerical result even further. This
suggests that the DS equations can be used to provide extremely accurate
solutions for the Green's functions, even when $D>0$, but these equations must
be supplemented by including the large-$n$ asymptotic behavior of the Green's
function $G_{2n}$. This asymptotic behavior cannot be determined from the DS
equations; it must be obtained from a large-$n$ asymptotic approximation to the
integral representing the Green's function.

\section{$D=0$ non-Hermitian cubic theory}\label{s5}
This section considers the cubic massless non-Hermitian $\cPT$-symmetric
Lagrangian
\begin{equation}
\cL=\third ig\phi^3.
\label{e5.1}
\end{equation}
For (\ref{e5.1}) the connected one-point Green's function is
\begin{equation}
G_1=\int dx\,x\exp(-ix^3/3)\Big/\int dx\,\exp(-ix^3/3),
\label{e5.2}
\end{equation}
where we take $g=1$. The path of integration lies inside a $\cPT$-symmetric pair
pair of Stokes sectors. These integrals can be evaluated exactly:
\begin{equation}
G_1=-i3^{1/3}\Gamma\big(\tthird\big)/\Gamma\big(\third\big)=-0.729\,011\,13...
\,i.
\label{e5.3}
\end{equation}

The DS equations for the Lagrangian (\ref{e5.1}) are simpler than those in
(\ref{e4.4}) for the Hermitian quartic theory. The first 19 DS equations are
given by
\begin{eqnarray}
G_2 \!&=& \!-G_1^2,\nonumber\\
G_3 \!&=& \!-2G_1G_2-i,\nonumber\\
G_4 \!&=& \!-2G_2^2-2G_1G_3,\nonumber\\
G_5 \!&=& \!-6G_2G_3-2G_1G_4,\nonumber\\
G_6 \!&=& \!-6G_3^2-8G_2G_4-2G_1G_5,\nonumber\\
G_7 \!&=& \!-20G_3G_4-10 G_2G_5-2G_1G_6,\nonumber\\
G_8 \!&=& \!-20G_4^2-30G_3G_5-12G_2G_6-2G_1G_7,\nonumber\\
G_9 \!&=& \!-70G_4G_5-42G_3G_6-14G_2G_7-2G_1G_8,\nonumber\\
G_{10}\!&=& \!-70G_5^2-112G_4G_6-56G_3G_7-16G_2G_8\nonumber\\
\!&&\!-2G_1G_9,\nonumber\\
G_{11}\!&=& \!-252G_5G_6 -168G_4G_7-72G_3G_8\nonumber\\
\!&&\!-18G_2G_9 -2G_1G_{10},\nonumber\\
G_{12}\!&=&\!-252G_6^2-420G_5G_7-240G_4G_8\nonumber\\
\!&&\!-90G_3G_9-20G_2G_{10}-2G_1G_{11},\nonumber\\
G_{13}\!&=& \!-924G_6G_7-660G_5G_8-330G_4G_9-110G_3G_{10}\nonumber\\
\!&&\!-22G_2G_{11} -2G_1G_{12},\nonumber\\
G_{14}\!&=&\!-924G_7^2-1584G_6G_8-990G_5G_9-440G_4G_{10}\nonumber\\
\!&&\!-132G_3G_{11}-24G_2G_{12}-2G_1G_{13},\nonumber\\
G_{15}\!&=& \!-3432G_7G_8-2574G_6G_9-1430G_5G_{10}\nonumber\\
\!&&\!-572G_4G_{11}-156G_3G_{12}-26G_2G_{13}-2G_1G_{14},\nonumber\\
G_{16}\!&=& \!-3432G_8^2 - 6006 G_7G_9 - 4004G_6G_{10}\nonumber\\
\!&&\!-2002G_5G_{11}-728G_4G_{12}-182G_3G_{13}\nonumber\\
\!&& \!-28G_2G_{14}-2G_1G_{15},\nonumber\\
G_{17}\!&=& \!-12870G_8G_9-10010G_7G_{10}-6006G_6G_{11}\nonumber\\
\!&&\!-2730G_5G_{12}-910G_4G_{13}\nonumber\\
\!&& \!-210G_3G_{14} -30G_2G_{15} -2G_1G_{16},\nonumber\\
G_{18}\!&=&\!-12870G_9^2-22880G_8G_{10}-16016G_7G_{11}\nonumber\\
\!&&\!-8736G_6G_{12}-3640G_5G_{13}-1120G_4G_{14}\nonumber\\
\!&& \!-240G_3G_{15} -32G_2G_{16} -2G_1G_{17},\nonumber\\
G_{19}\!&=& \!-48620G_9G_{10} -38896G_8G_{11}-24752G_7G_{12}\nonumber\\
\!&& \!-12376G_6G_{13} -4760G_5G_{14} -1360G_4G_{15}\nonumber\\
\!&& \!-272G_3G_{16}-34G_2G_{17}-2G_1G_{18},\nonumber\\
G_{20}\!&=&\!-48620G_{10}^2-87516G_9G_{11}-63648G_8G_{12}
\label{e5.4}\\
\!&&\!-37128G_7G_{13}-17136G_6G_{14} -6120G_5G_{15}\nonumber\\
\!&&\!-1632G_4G_{16}-306G_3G_{17}-36G_2G_{18}-2G_1G_{19}.\nonumber
\end{eqnarray}
The coefficients in these equations can be checked easily; the sum of the
coefficients on the right side of each equation is an increasing power of 2. For
example, for $G_8$ the sum of the coefficients is $20+30+12+2=2^6$, and for
$G_9$ the sum is $70+42+14+2=2^7$.

As in Sec.~\ref{s4}, we again use the unbiased truncation scheme of setting
higher-order Green's functions to zero. We obtain the leading approximation to
$G_1$ by substituting the first of these equations into the second and
truncating by setting $G_3=G_4=\dots=0$. The resulting cubic equation $G_1^3=
\half i$ has three solutions, and we choose the solution that is consistent
with $\cPT$ symmetry:
\begin{equation}
G_1=-2^{-1/3}i=-0.793\,700\,53...\,i.
\label{E5.5}
\end{equation}
This result differs by $8.9\%$ from the exact value of $G_1$ in (\ref{e5.3}).
However, the accuracy improves if we include more DS equations: We close the
system by using the first equation to eliminate $G_2$, the second to eliminate
$G_3$, and so on. The result is that the right side of the $G_n$ equation
becomes a polynomial of degree $n$ in the variable $G_1$, and we truncate the
system by setting the left side to zero and finding the roots of this
polynomial.

At first, the roots consistent with $\cPT$ symmetry that are obtained with this
procedure seem to approach the exact value of $G_1$ in (\ref{e5.3}) but unlike
the roots for the Hermitian quartic theory, where the approach is monotone (see
Fig.~\ref{F10}), the approach here is oscillatory at first: For the $n=4$
truncation the closest root is $-0.693\,361\,27...\,i$, which is smaller in
magnitude than the exact value of $G_1$, and for $n=5$ the closest root is
$-0.746\,900\,79...\,i$, which is larger in magnitude than the exact value. This
pattern seems to persist: For $n=6$ the closest root is $-0.712\,564\,55...\,i$
and for $n=7$ the closest root is $-0.739\,871\,08...\,i$. However, for $n=8$
this pattern breaks: The closest root is $-0.712\,368\,70...\,i$, which is
smaller in magnitude than the exact value, but is a slightly worse approximation
than the $n=6$ root.

The departure from the oscillatory convergence pattern at $n=8$ signals a new
behavior. The closest root for $n=9$ is $G_1=-0.738\,595\,46...\,i$, which is
slightly better than the $n=7$ root, but for $n=10$ we observe a qualitative
change in the character of the approximants. The polynomial associated with
$G_{10}$ is
$$G_{10}=40\big(9072 G_1^{10}-7560iG_1^7-1881G_1^4+119 iG_1\big).$$
If we truncate by setting the right side to zero and ignore the trivial root at
0, we see that all nontrivial roots come in triplets located at the vertices of
equilateral triangles. The roots that are closest to the exact value of $G_1$,
which lies on the negative-imaginary axis, are {\it not pure imaginary}. Rather,
there is a pair of roots close to and on {\it either side of the
negative-imaginary axis} at $-0.717\,367\,67...\,i\pm 0.016\,050\,677...\,.$

\begin{figure}[t]
\centering
\includegraphics[scale = 0.19]{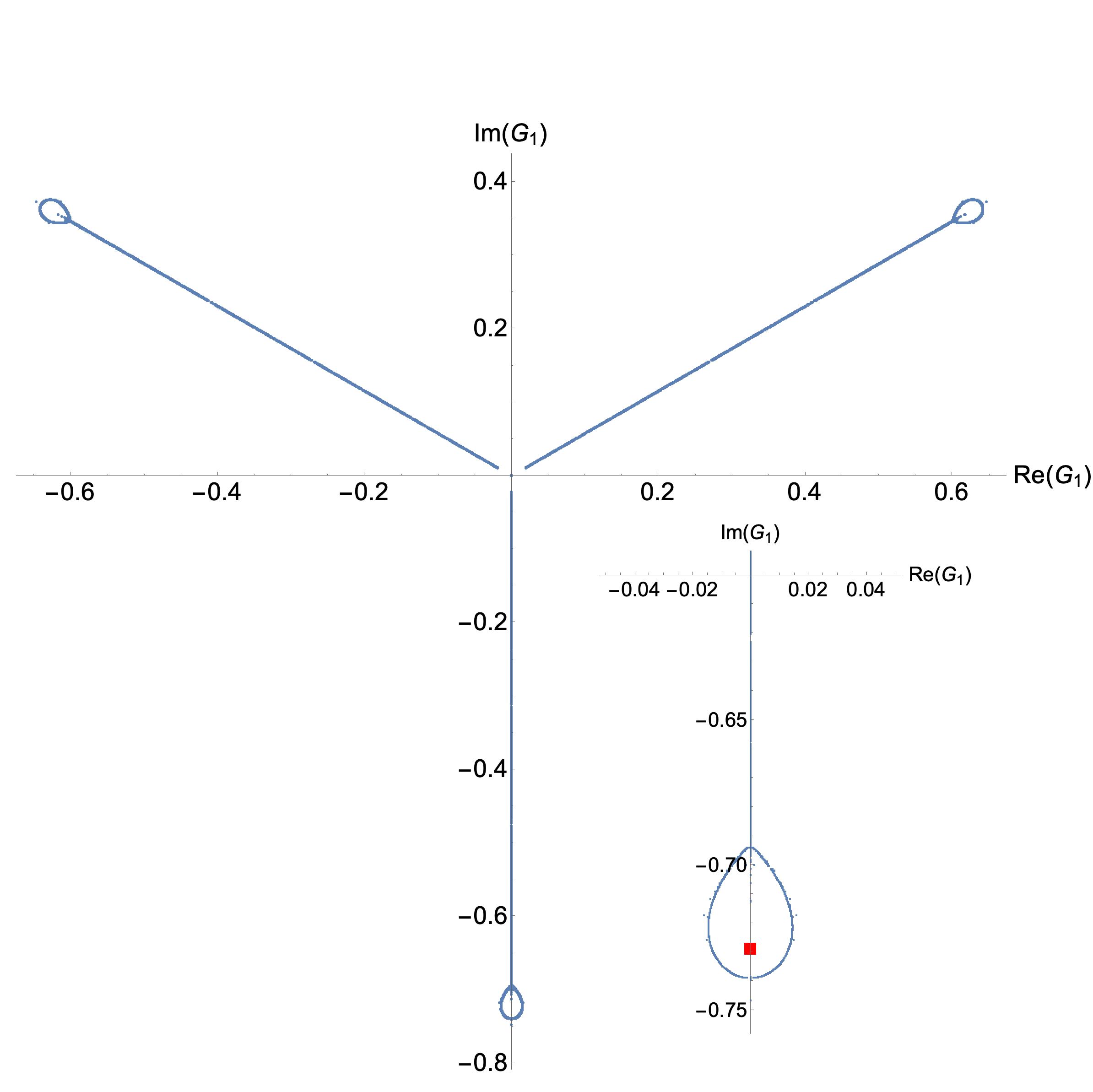}
\caption{All solutions $G_1$ to the DS equations from the third to the 200th
truncation plotted in the complex plane. The exact value of $G_1$ is
$-0.72901113...\,i$. The full set of solutions has threefold symmetry; solutions
lie on a three-bladed propeller with each blade having a small dense loop of
solutions at the end. The inset shows that the exact solution (red square) lies
on the negative-imaginary axis inside this loop.}
\label{F12}
\end{figure}

For higher truncations we find an accumulation of roots near the exact
negative-imaginary value in (\ref{e5.3}), but {\it arranged in a ring around
this exact value}. We have solved the DS equations up to the 200th truncation
and we plot the solutions as dots in the complex plane in Fig.~\ref{F12}.

\begin{figure}[t]
\centering
\includegraphics[scale = 0.2]{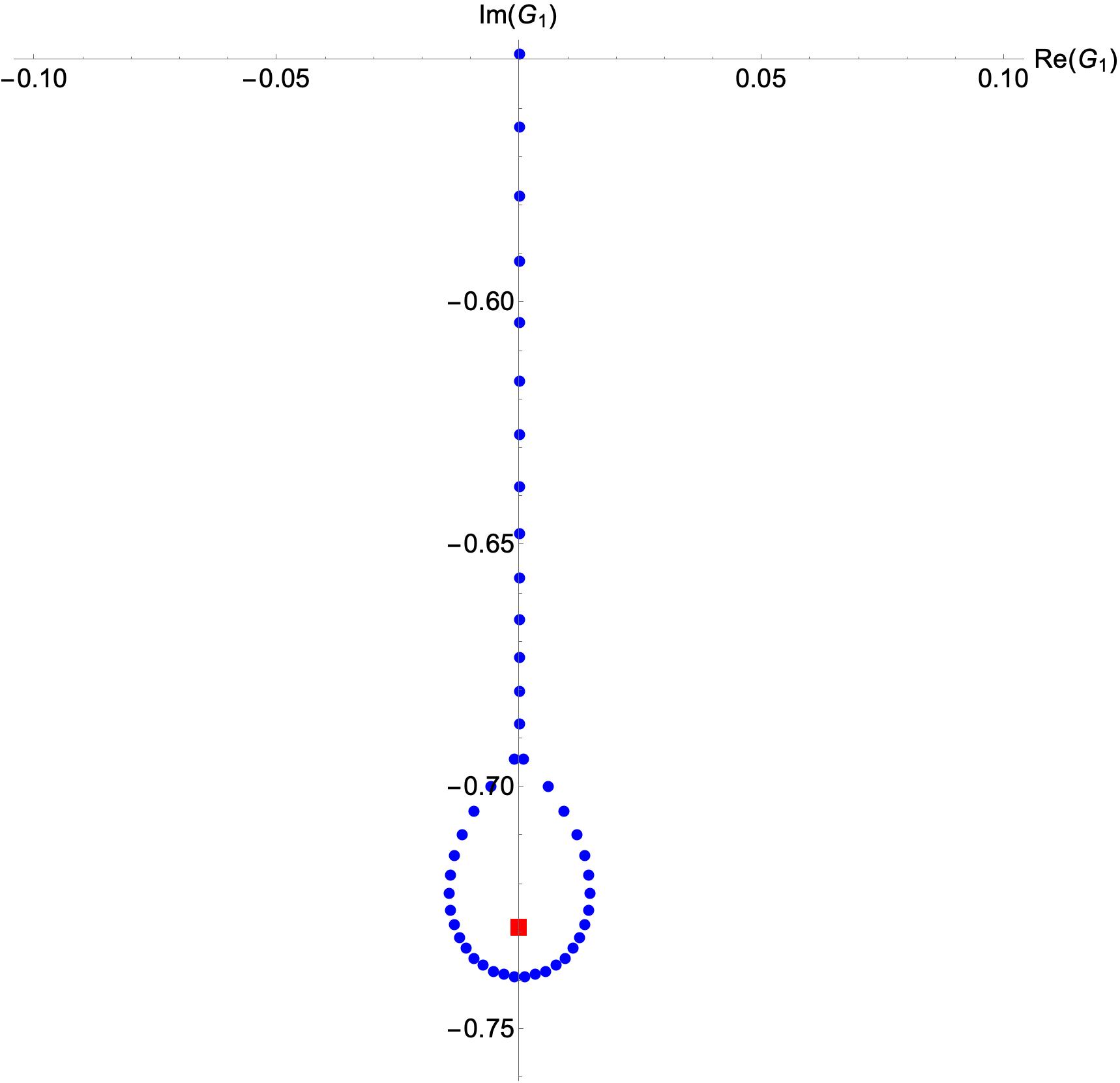}
\caption{[color online] Blowup of the negative imaginary axis for the solution
$n=200$ showing the loop-shaped concentration of solutions for $G_1$ around the
exact value of $G_1$, which lies in the interior of the loop as shown in
Fig.~\ref{F12}.}
\label{F13}
\end{figure}

We seek solutions that are near the negative-imaginary axis for two reasons:
First, $\cPT$ symmetry requires that $G_1$ be negative imaginary. Second, the
first equation in (\ref{e5.4}), $G_2=-G_1^2$, shows that otherwise $G_2$ will
not be positive; the second Green's function must be positive because $G_2=M^{
-2}$, where $M$ is the renormalized mass. A blow-up of the ring structure on the
negative-imaginary axis is shown in Fig.~\ref{F13} for the solution to the
$n=200$ polynomial only. This emphasizes that the roots on the ring are not
approaching the exact value of $G_1$ shown in Fig.~\ref{F12} as $n$ increases,
but rather are just becoming dense on the ring.

The three-fold symmetry of the roots in Fig.~\ref{F12} arises because the monic
polynomial equations that come from solving successively truncated DS equations
contain only powers of $x^3$ (after we exclude the trivial roots at 0):
\begin{equation}
P_{3n}(x)=x^{3n}+C_1 x^{3n-3}+C_2 x^{3n-6}+... +C_n.
\label{e5.7}
\end{equation}
Five such polynomials (with factors of $i$ excluded) are
\begin{eqnarray}
P_3 &=& x^3+\tfrac{1}{2},\nonumber\\
P_6 &=& x^6+\tfrac{1}{2}x^3+\tfrac{1}{20},\nonumber\\
P_9 &=& x^9+\tfrac{3}{4}x^6+\tfrac{87}{560}x^3+\tfrac{1}{160},\label{e5.8}
\\
P_{12} &=& x^{12}+x^9+\tfrac{93}{280}x^6+\tfrac{13}{336}x^3+\tfrac{7}{8800},
\nonumber\\
P_{15} &=& x^{15}+\tfrac{5}{4}x^{12}+\tfrac{4}{7}x^9+\tfrac{19}{168}x^6+
\tfrac{56909}{6726720}x^3+\tfrac{1}{9856}.\nonumber
\end{eqnarray}
Like the coefficients of the polynomials in (\ref{e4.5}), the coefficients $C_k$
in (\ref{e5.7}) have a fairly simple structure:
\begin{eqnarray}
C_1(n)&=& \tfrac{1}{1!}\big(\fourth\big)^1 n\quad(n>1),\nonumber\\
C_2(n)&=& \tfrac{1}{2!}\big(\fourth\big)^2\big(n^2-
\tfrac{47}{35}n\big)\quad(n>2),\nonumber\\
C_3(n)&=& \tfrac{1}{3!}\big(\fourth\big)^3 \big(n^3 -\tfrac{47\cdot
3}{35}n^2+\tfrac{134}{35}n\big)\quad(n>3),\nonumber\\
C_4(n)&=& \tfrac{1}{4!}\big(\fourth\big)^4
\big(n^4-\tfrac{47\cdot 6}{35}n^3
+\tfrac{25387}{1225}n^2
-\tfrac{1471121}{175175}n\big) \nonumber \\
&& \qquad(n>4).
\label{e5.9}
\end{eqnarray}

\subsection{Asymptotic behavior of $G_n$ for large $n$}
In Sec.~\ref{s4} we investigated the large-$n$ asymptotic behavior of the
Green's functions in order to study the validity of the truncation procedure for
the quartic Hermitian theory. We repeat this analysis for the non-Hermitian
cubic theory. The DS equations (\ref{e5.3}) and (\ref{e5.4}) determine the exact
values of the $G_n$. These are listed in Table \ref{t2}.

\begin{table}[h!]
\caption{Exact values of the first 14 nonzero connected Green's functions
for the $\cPT$-symmetric cubic theory.}
\centering
\begin{tabular}{|c|c|}
\hline\hline
$G_2^{\rm exact} = 0.531\,457\,23 $&
$G_3^{\rm exact} = -0.225\,123\,53\, i $\\
$G_4^{\rm exact} = -0.236\,658\,45$  &
$G_5^{\rm exact} = 0.372\,807\,88 \,i$\\
$G_6^{\rm exact} = 0.766\,712\,18$  &
$G_7^{\rm exact} = -1.928\,978\,72 \,i$\\
$G_8^{\rm exact} = -5.715\,182\,10 $ &
$G_9^{\rm exact} = 19.444\,890\,40 \,i$\\
$G_{10}^{\rm exact} = 74.616\,669\,21$ &
$G_{11}^{\rm exact} = -318.582\,603\,45 \,i$\\
$G_{12}^{\rm exact} = -1,497.372\,869\,48$ &
$G_{13}^{\rm exact} = 7,680.861\,833\,65 \,i$\\
$G_{14}^{\rm exact} = 42,692.806\,116\,42$  &
$G_{15}^{\rm exact} = -255,589.034\,701\,83 \,i$\\
\hline
\end{tabular}
\label{t2}
\end{table}

Applying Richardson extrapolation to the entries in Table \ref{t2}, we find
that the asymptotic behavior of $G_n$ for large $n$ (including the overall
multiplicative constant) is
\begin{equation}
G_n\sim-(n-1)!\,r^n(-i)^n\quad(n\to\infty),
\label{e5.10}
\end{equation}
where $r=0.427\,696\,347\,707...\,.$

This asymptotic behavior is confirmed analytically in Ref.~\cite{R1}. The
derivation goes as follows. We define
$$g_p\equiv-i^n G_p/(p-1)!$$
and express the DS equations for the Green's functions $G_{n}$ in compact form
as a recursion relation:
$$g_p=\frac{1}{p-1}\sum_{k=1}^{p-1}g_k g_{p-k}+\frac{1}{2}\delta_{p,3}\qquad
(p\geq2).$$
We then multiply by $(p-1)x^p$ to get $x^p (p-1) g_p =\sum_{k=1}^{p-1}g_k x^k
g_{p-k} x^{p-k}+x^3\delta_{p,3}$, and rewrite the left side as $x\frac{d}{dx}
x^p g_p-x^p g_p$. Next, we sum in $p$ from 2 to $\infty$ and define the
generating function $f(x)$:
$$f(x)\equiv\sum_{p=1}^\infty x^p g_p.$$
This generating function satisfies the Riccati equation
$$xf'(x)-f(x)=f^2(x)+x^3.$$

We linearize this equation by substituting $f(x)=-xu'(x)/u(x)$ and
$$f'(x)=x\frac{[u'(x)]^2}{[u(x)]^2}-\frac{u'(x)}{u(x)}-x\frac{u''(x)}{u(x)}$$
into the Riccati equation. Four terms cancel and we get
$$u''(x)=-xu(x).$$

This is an Airy equation of negative argument whose general solution is $u(x)=
a{\rm Ai}(-x)+b{\rm Bi}(-x)$, where $a$ and $b$ are arbitrary constants. Thus,
\begin{equation}
f(x)=x\frac{a{\rm Ai}'(-x) + b{\rm Bi}'(-x)}{a{\rm Ai}(-x) + b{\rm Bi}(-x)}.
\label{e5.11}
\end{equation}
To determine the constants $a$ and $b$, we note that
$$f'(0)=g_1=-3^{1/3}\Gamma\big(\tthird\big)/\Gamma\big(\third\big)=
-0.729\,011\,132\,947...\,.$$
Hence,
$$-3^{1/3}\Gamma\big(\tthird\big)/\Gamma\big(\third\big)=
\frac {a{\rm Ai}'(0) + b{\rm Bi}'(0)}{a{\rm Ai}(0) + b{\rm Bi}(0)}$$
We then substitute
\begin{eqnarray}
{\rm Ai}(0)&=&3^{-2/3}/ \Gamma\big(\tthird\big),~
{\rm Ai}'(0)=-3^{-1/3}/ \Gamma\big(\third\big),\nonumber\\
{\rm Bi}(0)&=&3^{-1/6}/ \Gamma\big(\tthird\big),~
{\rm Bi}'(0)=3^{1/6}/ \Gamma\big(\third\big),\nonumber
\end{eqnarray}
cancel the Gamma functions, and obtain $-1=(-a+b\sqrt{3})/(a+b\sqrt{3})$.
Thus, $a$ is arbitrary and $b=0$, so
$$f(x) = x{\rm Ai}'(-x)\big/{\rm Ai}(-x).$$

The generating function $f(x)$ is a power series, and it blows up when the
denominator in this equation is zero. This happens first when
$x=2.338\,107\,410\,459...\,$, which is the radius of convergence of the series.
The {\it inverse} of this number is precisely the value of $r$ in (\ref{e5.10}).

\begin{figure}[t!]
\centering
\includegraphics[scale = 0.20]{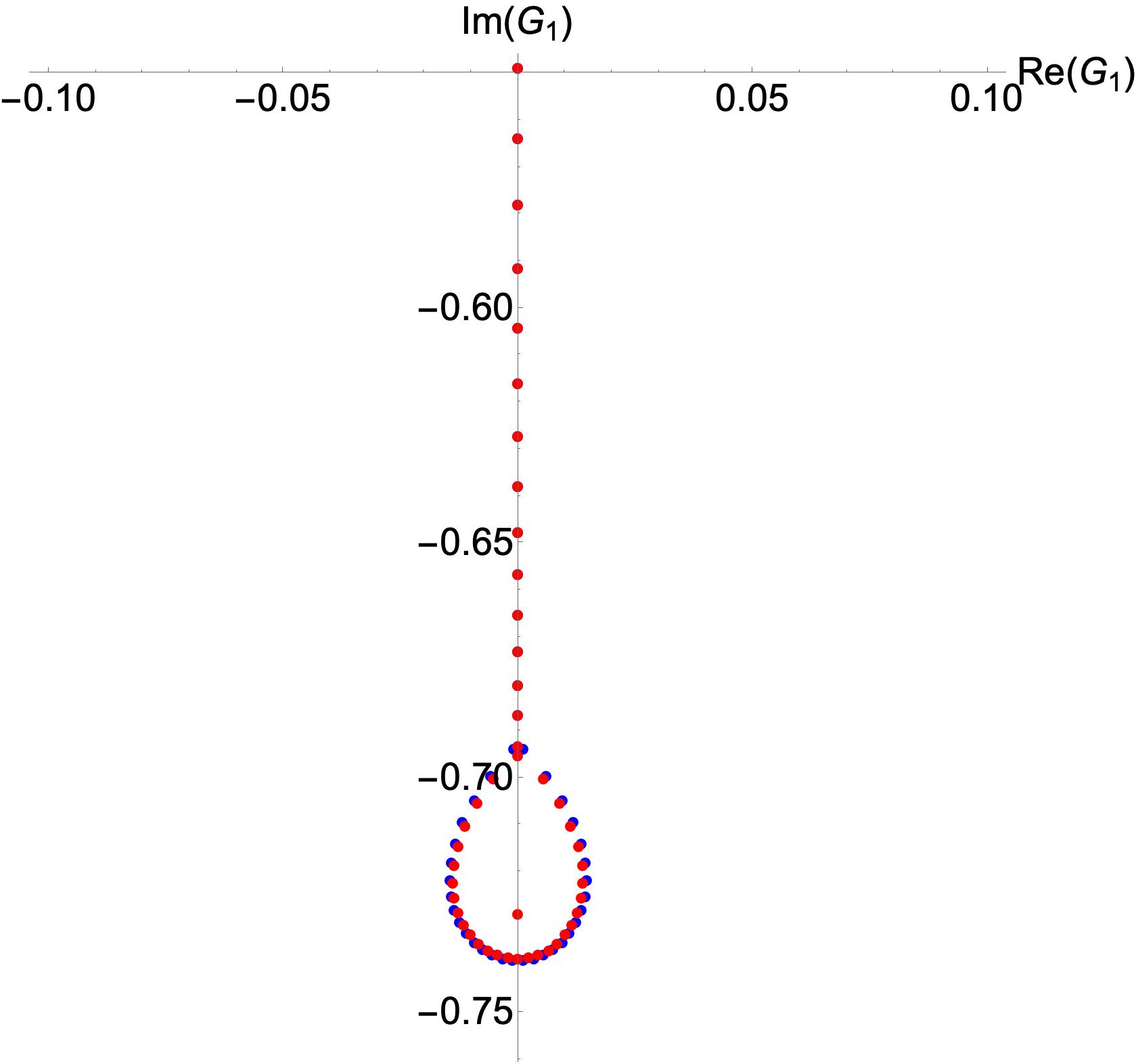}
\caption{[color online] Blowup of the negative imaginary axis for the $n=200$
truncation showing solutions for $G_1$, obtained from (i) the unbiased (blue)
and (ii) the asymptotic approximations (red) for $G_n$. The blue and red
loops are almost the same size, but there is a new red dot that is almost
exactly equal to $G_1$. These dots also appear in the two other loops at the
ends of the three-bladed propeller.}
\label{F14}
\end{figure}

Once again, we are faced with justifying the truncation needed to solve the
system of DS equations and we repeat the argument in Sec.~\ref{s4}. As before,
the unbiased truncation gives a slowly converging sequence of approximants that
does not converge to the exact value of $G_1$. The novelty here is that, if we
use the asymptotic expression (\ref{e5.10}) as the basis of the truncation, an
entirely new root, which is extremely close to the exact value of $G_1$, appears
inside the tight loop of roots in the complex plane, as shown in Fig.~\ref{F14}.
This figure gives a comparison of the $n=200$ evaluation using this asymptotic
approximation (red) and the unbiased truncation (blue). The blue and red loops
are almost the same size, but the new root agrees with the exact value of $G_1$
to seven decimal places. However, corresponding new roots also appear in the
loops at the ends of the other two propellers. The condition of global $\cPT$
symmetry does not exclude these roots because the entire constellation of zeros
is $\cPT$ symmetric. To exclude these spurious zeros we can impose the condition
that the $G_2$ be positive (spectral positivity). We do so by using the first DS
equation in (\ref{e5.4}).

\begin{figure}[t!]
\centering
\includegraphics[scale = 0.29]{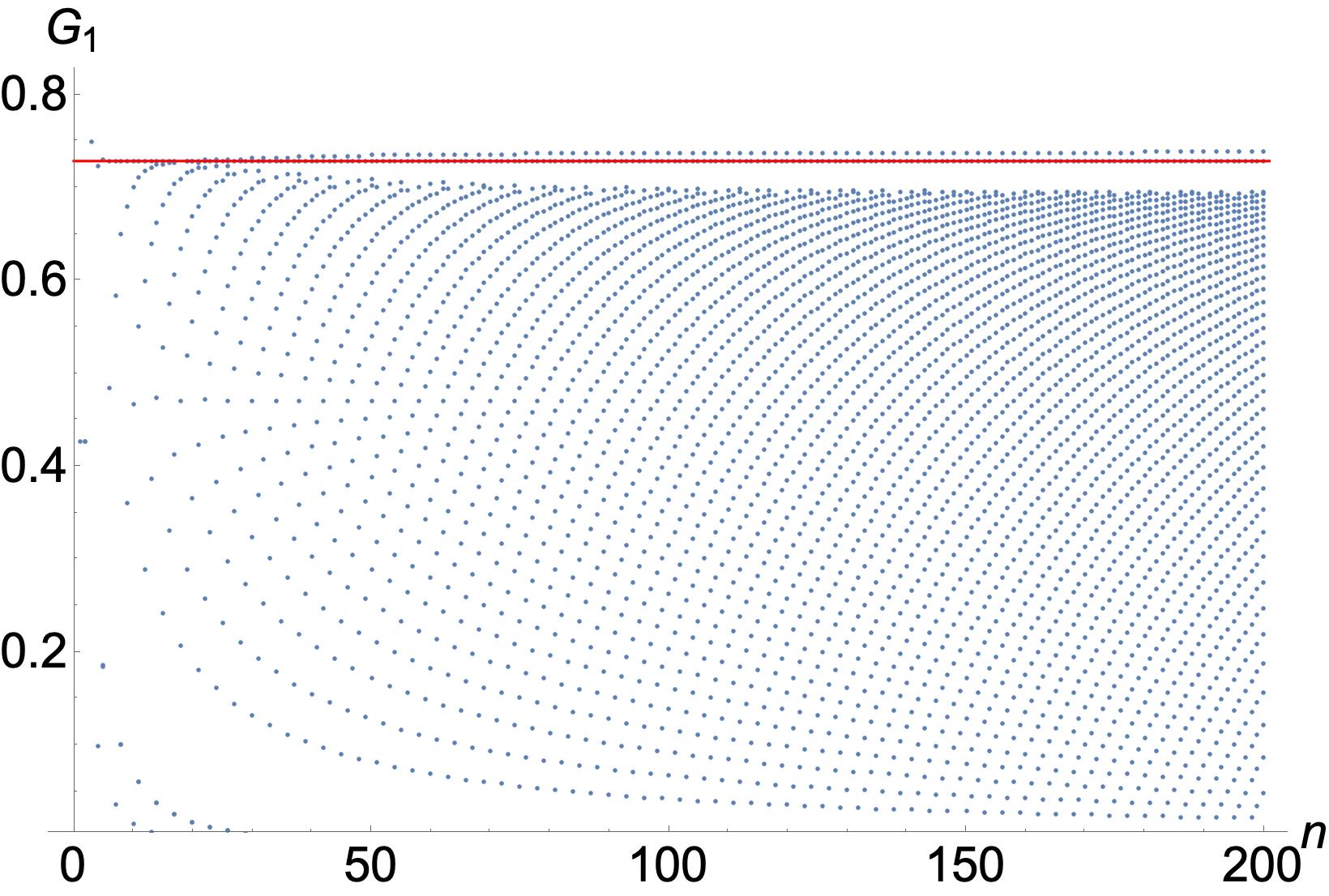}
\caption{[color online] Absolute values of the solutions for $G_1$ on the
imaginary axis, obtained using the asymptotic approximation to $G_n$ for $n$
ranging from 1 up to 200. The heavy red line shows the exact value of
$\big|G_1\big|$.}
\label{F15}
\end{figure}

To see more clearly the effect of including the asymptotic behavior of $G_n$ in
the truncation scheme, we plot the absolute values of the solutions along the
negative axis for $n$ ranging from 1 to 200 in Fig.~\ref{F15}. As we see in
Fig.~\ref{F14}, there are solutions which are both larger and smaller (in
absolute value) than the exact solution. Thus, we do not observe a monotonic
behavior of the roots for increasing $n$. However, the the isolated root inside
the loop in Fig.~\ref{F14} is indistinguishable from the exact solution (red
line).

\section{$D=0$ non-Hermitian quartic theory}\label{s6}
To understand more broadly the behavior of our truncation schemes, we consider
next the quartic Lagrangian
\begin{equation}
\cL=-\fourth g\phi^4,
\label{e6.1}
\end{equation}
which defines a non-Hermitian massless $\cPT$-symmetric theory in
zero-dimensional spacetime. The connected one-point Green's function is for this
Lagrangian
\begin{equation}
G_1=\int dx\,x\exp(gx^4/4)\Big/\int dx\,\exp(gx^4/4),
\label{e6.2}
\end{equation}
where the paths of integration lie inside a $\cPT$-symmetric pair of Stokes
sectors of angular opening $\tfrac{\pi}{4}$ centered about $-\tfrac{\pi}{4}$
and $-\tfrac{3\pi}{4}$ in the lower-half complex-$x$ plane. Without loss of
generality, we set $g=1$ and evaluate these integrals exactly:
\begin{equation}
G_1=-2i\sqrt{\pi}\big/\Gamma\big(\fourth\big)=-0.977\,741\,07...\,i.
\label{e6.3}
\end{equation}

The first eight DS equations for this theory are
\begin{eqnarray}
G_3 &=& -G_1^3-3G_1G_2,\nonumber\\
G_4 &=& -3G_1G_3-3G_2^2-3G_1^2G_2-1,\nonumber\\
G_5 &=& -3G_1G_4-9G_2G_3-3G_1^2G_3-6G_1G_2^2,\nonumber\\
G_6 &=& -3G_1G_5-12G_2G_4-3G_1^2G_4-9G_3^2\nonumber\\
&& ~ -18G_1G_2G_3-6G_2^3,\nonumber\\
G_7 &=& -3G_1G_6-15G_2G_5-3G_1^2G_5-30G_3G_4\nonumber\\
&& ~ -24G_1G_2G_4-18G_1G_3^2-36G_2^2G_3,\nonumber\\
G_8 &=& -3G_1G_7-18G_2G_6-3G_1^2G_6-45G_3G_5\nonumber\\
&& ~ -30G_1G_2G_5-30G_4^2-60G_1G_3G_4\nonumber\\
&& ~ -60G_2^2G_4-90G_2G_3^2,\nonumber\\
G_9 &=& -3G_1G_8-21G_2G_7-3G_1^2G_7-63G_3G_6\nonumber\\
&& ~ -36G_1G_2G_6-105G_4G_5-90G_1G_3G_5\nonumber\\
&& ~ -90G_2^2G_5-60G_1G_4^2-360G_2G_3G_4-90G_3^3,\nonumber\\
G_{10} &=& -3G_1G_9-24G_2G_8-3G_1^2G_8-84G_3G_7\nonumber\\
&& ~ -42G_1G_2G_7-168G_4G_6-126G_1G_3G_6\nonumber\\
&& ~ -126G_2^2G_6-105G_5^2-210G_1G_4G_5\nonumber\\
&& ~ -630G_2G_3G_5-420G_2G_4^2-630G_3^2G_4.
\label{e6.4}
\end{eqnarray}

The unbiased approach to solving these equations consists of fixing $n$ and then
using successive linear elimination to obtain polynomial equations to be solved
numerically for the lowest Green's functions. However, the procedure is more
difficult than for the Hermitian quartic theory in (\ref{e4.1}) or the
non-Hermitian cubic theory in (\ref{e5.1}) because this elimination process
concludes with {\it two} polynomials containing not one but two Green's
functions $G_1$ and $G_2$. That is, we obtain a {\it coupled pair} of polynomial
equations to solve for $G_1$ and $G_2$ rather than one polynomial equation in
one Green's function.

For example, the leading truncation ($n=4$) consists of eliminating $G_3$ in
the second DS equation by substituting the first DS equation into it. We then
truncate by setting $G_3=G_4=G_5 =\dots=0$ and solve the resulting {\it pair}
of simultaneous equations. This leads to $G_1^4=3/2$, and the $\cPT$-symmetric
solution in the negative-half plane is
\begin{equation}
G_1=-i\,\big(\threehalf\big)^{1/4}=-1.106\,681\,92...\,i.
\label{e6.5}
\end{equation}
This result has an error of $13.2\%$ in comparison with the exact value of
$G_1$ in (\ref{e6.3}).

For larger values of $n$ the procedure for solving the pair of polynomial
equations is tedious: We multiply each equation by an expression that makes the
coefficient of highest power of $G_1$ (or $G_2$) the same and then subtract the
two equations to eliminate this highest-power term. We repeat this process until
one of the equations becomes {\it linear} in $G_2$. We solve this equation for
$G_2$ and eliminate it algebraically from the other equation. This gives a
high-degree polynomial equation for $G_1$ that we can finally solve numerically.

The problem with this procedure is that each multiplication introduces spurious
roots. However, we find that the final polynomial in powers of $G_1$ {\it
factors into two polynomials}; the roots of one factor are all spurious while
the roots of the other factor, which is a polynomial in powers of $G_1^4$, solve
the original pair of equations. The number of roots increases rapidly with $n$
and all roots come in quartets that lie at the vertices of squares in the
complex plane. All (nonspurious) roots up to $n=33$ are displayed in
Fig.~\ref{F15}. The $\cPT$ symmetry of the Lagrangian (\ref{e6.1}) requires that
$G_1$ be a negative-imaginary number.

Since we must solve coupled equations for $G_1$ and $G_2$, we require the exact
value of $G_2$ in order to calculate the exact values of all of the Green's
functions from (\ref{e6.4}). The exact value of $G_2$ is
\begin{equation}
G_2=4\pi\big/\Gamma^2\big(\fourth\big)-2\Gamma\big(\threefourth\big)\big/\Gamma
\big(\fourth\big)=0.279\,999\,35...\,.
\label{e6.6}
\end{equation}
Then, using $G_1$ in (\ref{e6.3}) and $G_2$ in (\ref{e6.6}) we obtain the
results given in Table \ref{t3}.

\begin{table}[b!]
\caption{Exact values of 20 Green's functions for the
non-Hermitian quartic theory.}
\centering
\begin{tabular}{|c|c|}
\hline\hline
$G_3^{\rm exact} = -0.113\,397\,i$ &
$G_4^{\rm exact} = -0.099\,559$ \\
$G_5^{\rm exact} =  0.128\,446\,i$ &
$G_6^{\rm exact} =  0.215\,052\,$ \\
$G_7^{\rm exact} = -0.439\,386\,i$ &
$G_8^{\rm exact} = -1.055\,947$ \\
$G_9^{\rm exact} =  2.912\,307\,i$ &
$G_{10}^{\rm exact} = 9.055\,948$ \\
$G_{11}^{\rm exact} = -31.325\,429\,i$ &
$G_{12}^{\rm exact} = -119.269\,436$ \\
$G_{13}^{\rm exact} = 495.565\,822\,i$ &
$G_{14}^{\rm exact} = 2,231.100\,879$ \\
$G_{15}^{\rm exact} = - 10,818.525\,260\,i$ &
$G_{16}^{\rm exact} = - 56,209.003\,831$ \\
$G_{17}^{\rm exact} = 311,520.607\,892\,i$ &
$G_{18}^{\rm exact} = 1,834,444.674\,851$ \\
$G_{19}^{\rm exact} = - 11,438,011.031\,i$ &
$G_{20}^{\rm exact} = - 75,280,067.556$ \\
$G_{21}^{\rm exact} = 521,539,592.082\,i$ &
$G_{22}^{\rm exact} = 3,793,889,240.849$ \\
\hline
\end{tabular}
\label{t3}
\end{table}

\begin{figure}[h]
\centering
\includegraphics[scale = 0.27]{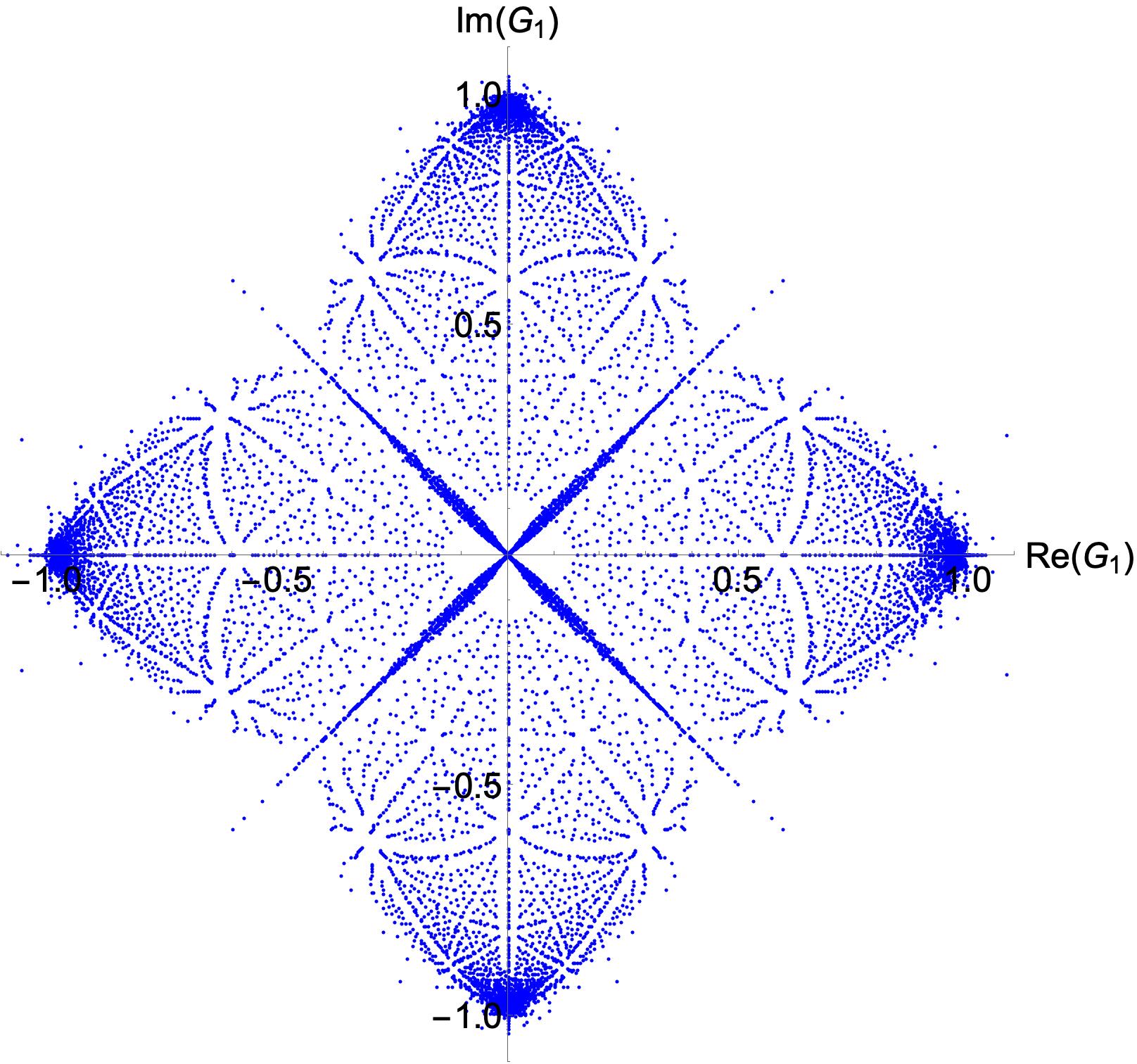}
\caption{[color online] All solutions for $G_1$ in the complex plane up to $n=
40$. The roots exhibit fourfold symmetry. However, $\cPT$ symmetry requires that
$G_1$ be negative imaginary, so only the roots on or near the negative-imaginary
axis are physically acceptable. The exact value of $G_1$, given in (\ref{e6.3}),
lies inside the concentration of roots near $-i$.}
\label{F16}
\end{figure}

\subsection{Asymptotic behavior of $G_n$ for large $n$}
Inspection of Table~\ref{t3} shows that the exact values of the odd (even)
Green's functions oscillate in sign as $n$ increases, and that the odd Green's
functions are imaginary, while the even ones are real. Applying Richardson
extrapolation to the entries in Table \ref{t3}, we find that the asymptotic
behavior of $G_n$ for large $n$ is
\begin{equation}
G_n\sim-(n-1)!\,(-i)^n r^n\quad(n\to\infty),
\label{e6.7}
\end{equation}
where $r=0.34640...$. (The overall multiplicative constant in the asymptotic
behavior is exactly 1.) This result is similar to the behavior in (\ref{e5.10})
for the Green's functions of the non-Hermitian cubic theory.

\section{$D=0$ non-Hermitian Quintic Theory} \label{s7}
Next, we analyze the quintic $\cPT$-symmetric Lagrangian
\begin{equation}
\cL=-\tfrac{1}{5}ig\phi^5.
\label{e7.1}
\end{equation}
The one-point Green's function is given by
\begin{equation}
G_1=\int dx\,x\exp(gx^5/5)\Big/\int dx\,\exp(gx^5/5).
\label{e7.2}
\end{equation}
Choosing $\cPT$-symmetric Stokes wedges in the negative half-plane
and setting $g=1$, we get the exact value 
\begin{equation}
G_1=-1.078\,653\dots\,.
\label{e7.3}
\end{equation}

\begin{figure}[h]
\centering
\includegraphics[scale = 0.27]{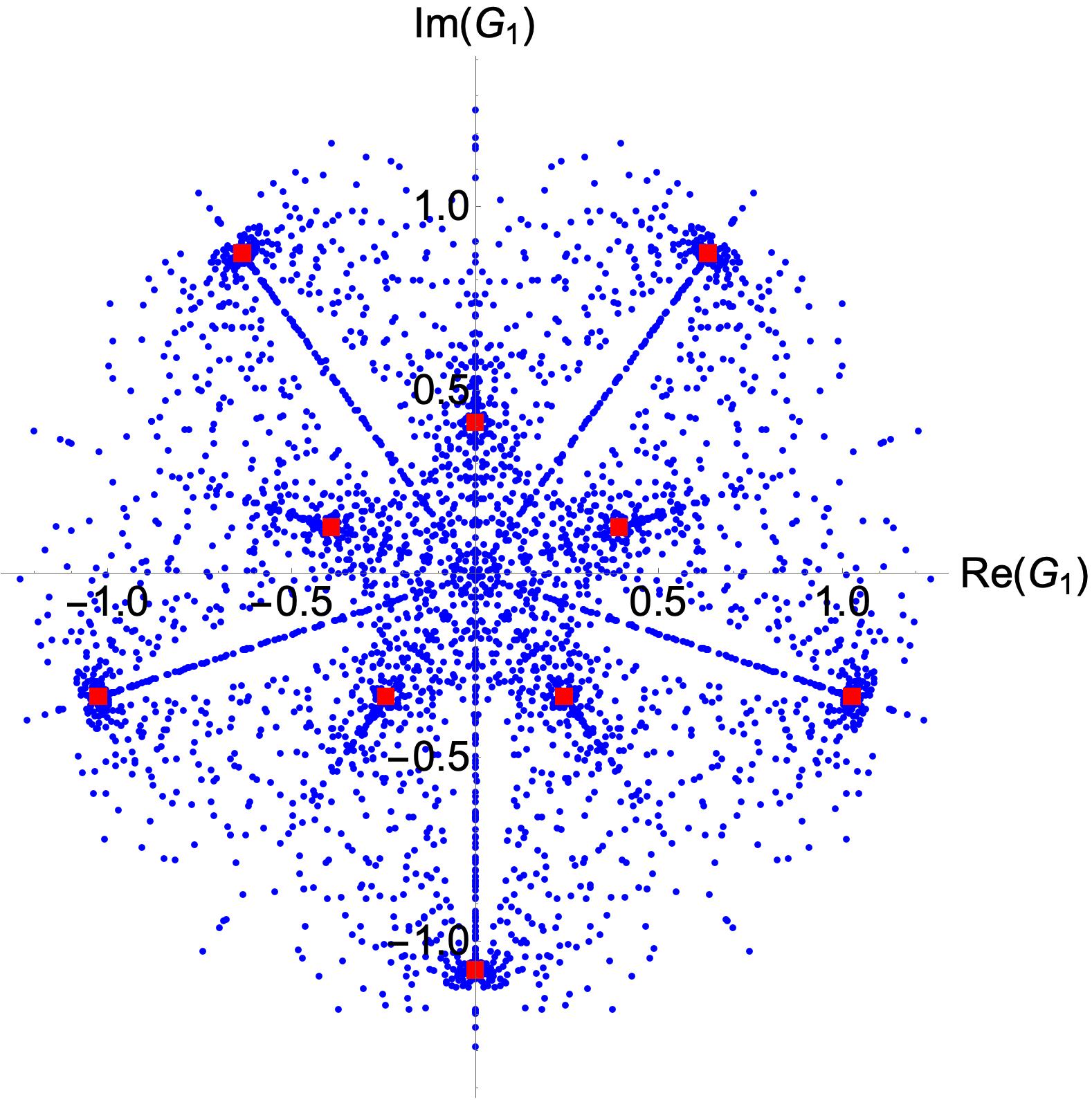}
\caption{[color online] All solutions for $G_1$ in the complex plane, up to
$n=11$. The roots exhibit fivefold symmetry but $\cPT$ symmetry requires that
$G_1$ be negative imaginary, so only the roots on or near the
negative-imaginary axis are physically acceptable. The exact value of $G_1$,
given in (\ref{e7.3}), lies inside the concentration of roots near $-i$.}
\label{F17}
\end{figure}

The first three DS equations that one obtains are
\begin{eqnarray}
G_4 &=& -G_1^4 -6G_2G_1^2 -4G_3G_1 - 3G_2^2 \nonumber \\
G_5 &=& -4G_2^3G_2 -12 G_1G_2^2 -6G_1^3G_3 -10G_2G_3 \nonumber \\
&&\quad-4G_1G_4 + i \nonumber \\
G_6 &=& -12 G_1^2G_2^2 -4 G_1^2G_3 - 12 G_2^3 - 36 G_1G_2G_3 \nonumber \\
&&\quad- 6 G_1^2G_4 - 10 G_3^2 - 14 G_2G_4 - 4 G_1G_5.
\label{e7.4}
\end{eqnarray}
The first equation for $G_4$ contains three unknowns, $G_1, G_2$ and $G_3$, so
setting $G_4=G_5=\dots=0$ as a first unbiased truncation means that we must
solve {\it three} coupled equations. At the next truncation $G_4\ne0$, but all
higher $G_n=0$. We therefore eliminate $G_4$ in terms of $G_1$, $G_2$, and
$G_3$, and must solve the next set of three equations. Thus, the solution is
complicated. Figure~\ref{F17} gives a plot of the roots in the complex plane up
to $n=11$. These roots exhibit five-fold symmetry.

We observe ten concentrations of roots. One can understand this as follows:
Associated with the Lagrangian (\ref{e7.1}) are five Stokes sectors that
define the regions of convergence of the integral (\ref{e7.2}) in the complex
plane (Stokes sectors). Thus, there are ten possible distinct paths of
integration in the complex plane, each of which lead to different values of
$G_1$. Aside from the imaginary value in (\ref{e7.3}), there is another
imaginary $\cPT$-symmetric solution from $\cPT$-symmetric (left-right symmetric)
integration in the upper-half plane that gives $G_1=0.4120\dots i$. The other
possible complex values of the integral for $G_1$ are $\pm 0.392\dots+0.127\dots
i$, $\pm 0.242\dots-0.333\dots i$, $\pm 0.634\dots+0.872\dots i$, and $\pm 1.025
\dots-0.333\dots i$. These ten values for $G_1$ are plotted as red squares on
Fig.~\ref{F17}, and correspond to the dense regions of solutions.

This feature is a general one: for the quartic and cubic systems discussed in
the previous sections, an analysis of all possible paths of integration in each
case results in {\it all} possible solutions of the Green's functions being
represented in the complex plane.

\section{$D=0$ Hermitian Sextic Theory}\label{s8}
Here we consider the zero-dimensional model described by the massless Lagrangian
\begin{equation}
\cL=\frac16 g\phi^6.
\label{e8.1}
\end{equation}
The first two Green's functions are given by 
\begin{eqnarray}
G_1 &=& \int d\phi \phi e^{-\phi^6/6}\Big/ \int d\phi e^{-\phi^6/6}\nonumber\\
G_2 &=& \int d\phi \phi^2e^{-\phi^6/6}\Big/ \int d\phi e^{-\phi^6/6}, 
\label{e8.2}
\end{eqnarray}
where $g=1$. The paths of integration must be specified, and there are six
distinct regions of convergence bounded by the Stokes lines at $\tfrac{\pi}{12}
+n\tfrac{\pi}{6}$, so there are 15 possible combinations of integration paths.
Thus, there are 15 different theories associated with the Lagrangian
(\ref{e8.1}).

The first DS equation is complicated and contains {\it five} unknowns:
\begin{eqnarray}
G_5 &=& -G_1^5-10G_1^3G_2-10G_1^2G_3-15G_1G_2^2,\nonumber\\
&& \qquad-5G_1G_4-10G_2G_3\,,
\label{e8.3}
\end{eqnarray}
and, in general, a complete solution of the DS equations would require that we
solve four coupled polynomial equations at each truncation level. 

To reduce the calculational complexity, we restrict the system to be parity
symmetric, so that all odd Green's functions vanish. The Hermitian sextic theory
has an integration path along the real axis. The exact value of $G_2$, obtained
by integrating (\ref{e8.2}) along this path gives
\begin{equation}
G_2=0.578\,616\,519\dots.
\label{e8.3x}
\end{equation}
However, there are two other choices for the integration path that also respect
parity symmetry and give rise to the values
\begin{equation}
G_2=-0.2893\dots\pm0.5010\dots i.
\label{e8.3y}
\end{equation}

Imposing parity symmetry on the first DS equation (\ref{e8.3}) gives the trivial
equation $0=0$. The first five (nontrivial) DS equations link the even-$n$
Green's functions to others of higher order: 
\begin{eqnarray}
G_6 &=& -15G_2^3-15G_2G_4+1,\nonumber\\
G_8 &=& -60G_2^4 -165G_2^2G_4 -35G_4^2-25G_2G_6,\nonumber\\
G_{10} &=& -120G_2^5-1200G_2^3G_4-1150G_2G_4^2-365G_2^2G_6\nonumber\\
&&-205 G_4G_6,\nonumber\\
G_{12} &=& -4200G_2^4G_4-16800G_2^2G_4^2-4550G_4^3\nonumber\\
&&-3360G_2^3G_6-8470G_2G_4G_6-455G_6^2\nonumber\\
&&-645G_2^2G_8-460G_4G_8-45G_2G_{10},\nonumber\\
G_{14} &=&-100800G_2^3G_4^2-168000G_2G_4^3-15120G_2^4G_6\nonumber\\
&&-151200G_2^2G_4G_6-76020G_4^2G_6-23310G_2G_6^2\nonumber\\
&&-7200G_2^3G_8-22680G_2G_4G_8-2910G_6G_8\nonumber\\
&&-1005G_2^2G_{10}-875G_4G_{10}-55G_2G_{12}.\nonumber
\end{eqnarray}

\begin{figure}[h]
\centering
\includegraphics[scale = 0.3]{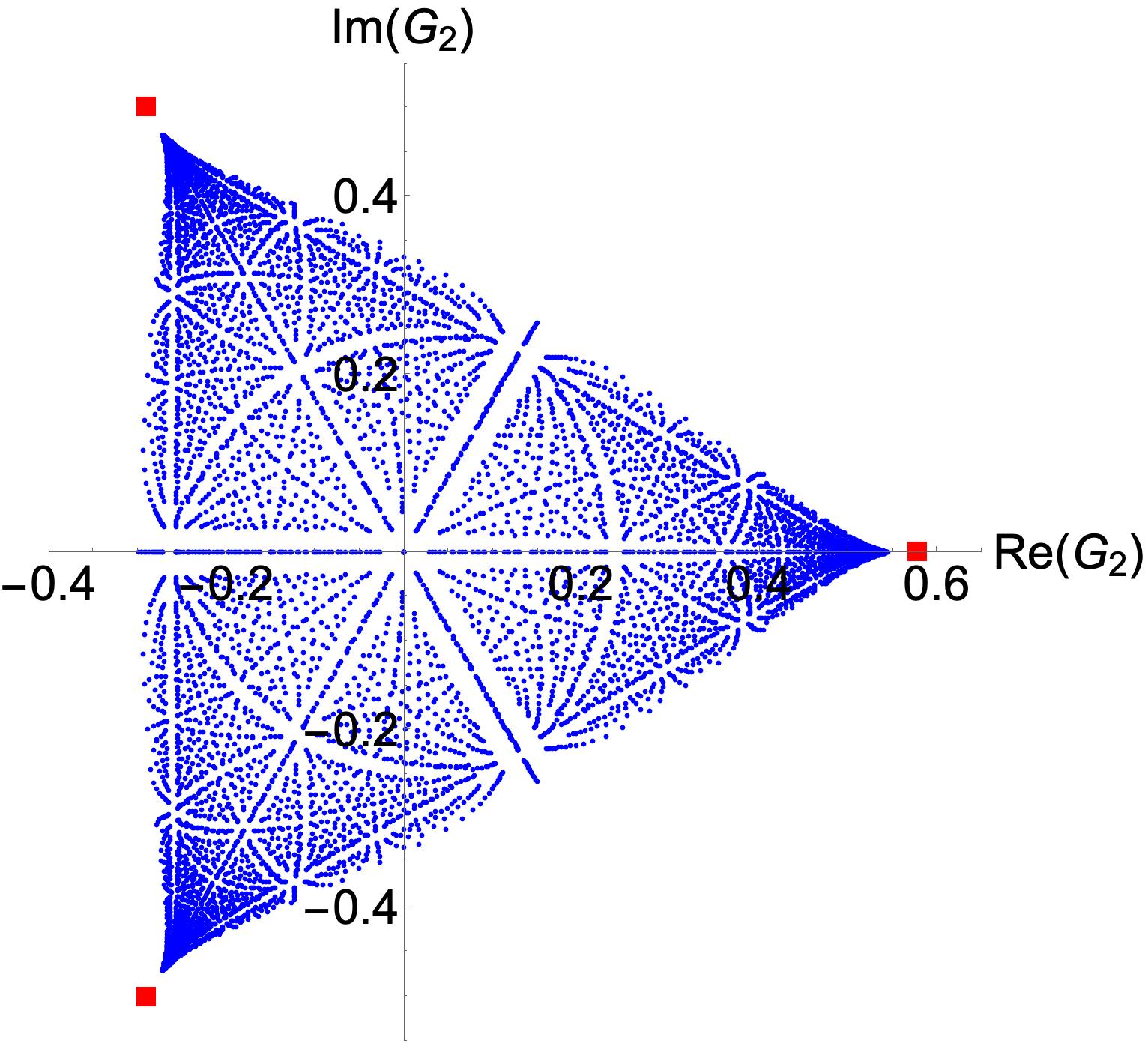}
\caption{[color online] Roots of the DS equations for the Hermitian sextic case 
obtained by means of the unbiased truncation scheme. There are three 
concentrations of solutions for $G_2$ that differ from the exact values in
(\ref{e8.3x}) and (\ref{e8.3y}) (red squares) by a few percent.}
\label{F18}
\end{figure}

As before, we truncate the DS equations by taking at each step a pair of
successive equations and setting the highest-order connected Green's functions
to zero. This constitutes a truncation of order $n$. The results for $G_2$ up to
$n=30$ (where we solve the equations $G_{64}=G_{66}=0$) are shown in
Fig.~\ref{F18}.

\begin{figure}[h]
\centering
\includegraphics[scale=0.25]{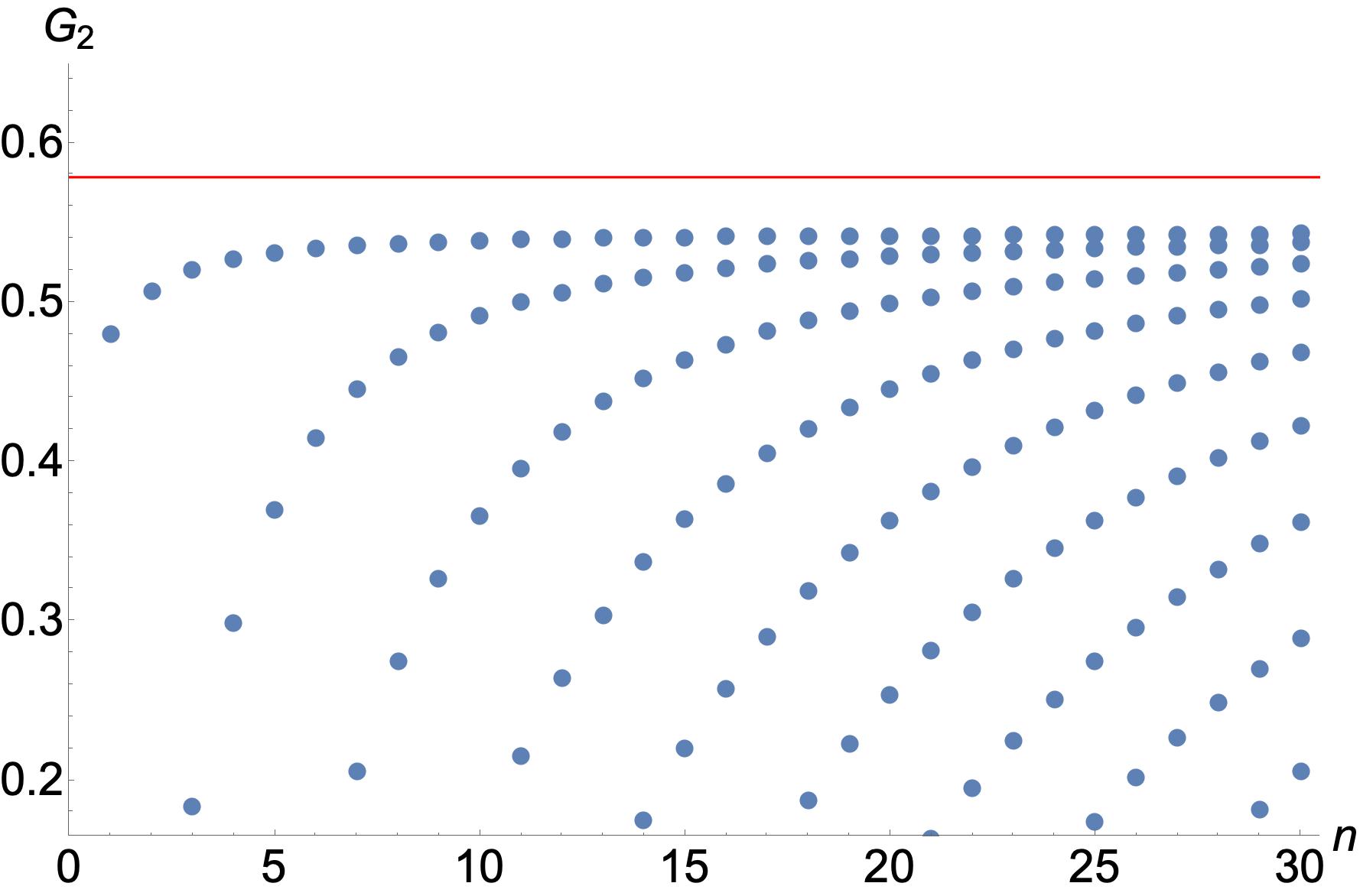}
\caption{[color online] Positive values of the roots of the SD equations for the
Hermitian sextic case (dots). The exact value of $G_2=0.578\,616\dots$ is
indicated by the red line and the DS results converge to a number that differs
from the exact value by about 20\%.}
\label{F19}
\end{figure}

Like the quartic case, the roots converge monotonically to points near the three
exact values of $G_2$ in (\ref{e8.3x}) and (\ref{e8.3y}). Richardson
extrapolation gives the limiting values of the truncated sequences and these
limiting values differ from the exact values by 6\% (see Fig.~\ref{F19}).

\section{Conclusions}\label{s9}
In this paper we have studied the effectiveness of the DS equations as a way
to calculate the Green's functions of a quantum field theory. We have examined
the DS equations for zero-dimensional field theories only because in this case
we can evaluate the integral representations of the Green's functions exactly
and then compare these exact results with the approximants provided by the DS
equations. We find that while the Green's functions exactly satisfy the infinite
system of coupled DS equations, the DS equations alone cannot be used to obtain
accurate results for the Green's functions.

The reason for this is that Green's functions are expressed in terms of moments
of the functional integral that specifies the partition function $Z$ of the
quantum field theory. However, the DS equations are derived by functional {\it
differentiation} of the partition function. While differentiation preserves {\it
local} information, the {\it global} information in the functional integral,
which is required to specify the Green's functions uniquely, is lost.

As a trivial example, consider the function
\begin{equation}
f(x)=\tfrac{1}{4}x^3+\tfrac{2}{x} \quad(1\leq x\leq2).
\label{e9.1}
\end{equation}
We may differentiate $f(x)$ once to obtain a differential equation satisfied by
$f(x)$:
\begin{equation}
f'(x)+\tfrac{f(x)}{x}=x^2.
\label{e9.2}
\end{equation}
However, while this equation describes the {\it local} behavior of $f(x)$ at
each point $x$, we have lost the global boundary data needed to recover the
original function $f(x)$: The general solution to this differential equation,
\begin{equation}
f(x)=\tfrac{1}{4}x^3+\tfrac{C}{x},
\label{e9.3}
\end{equation}
contains an arbitrary constant. However, if we specify the boundary data $f(2)=
3$, this determines that $C=2$ and we have recovered $f(x)$ in (\ref{e9.1}).

As explained in Sec.~\ref{s3}, deriving the DS equations involves a somewhat
more complicated differentiation process. However, the resulting coupled
infinite system of DS equations is so complicated that it obscures the simple
fact that in the differentiation process some information has been lost. For
example, the functional-integral representation of the partition function exists
because the path of functional integration terminates as $|\phi|\to\infty$
inside a pair of Stokes sectors in complex-$\phi$ space ($\phi$ is the
integration variable). Because there are many possible pairs of sectors that
give a convergent functional integral, when we solve the DS equations we find
{\it all possible} solutions to the DS equations corresponding to all possible
pairs of Stokes sectors, some corresponding to Hermitian theories and others
corresponding to non-Hermitian theories (both $\cPT$-symmetric and
non-$\cPT$-symmetric). For instance, in Fig.~\ref{F17} there are 10
concentrations of roots corresponding to the ten possible paths of integration
for a quintic field theory. The DS equations weight each of these theories
equally.

There is even more loss of information than this. As we have shown, solving the
DS equations is a two-step process. First, we truncate the infinite triangular
system of DS equations, but when we do so, the resulting finite system always
has more Green's functions than equations and is therefore indeterminate. Next,
to obtain a closed system of coupled equations we perform a further truncation
in which we set the highest Green's functions to 0. (In this paper we call this
truncation procedure unbiased.) There are other truncation possibilities as
well, but in all cases we find that as we include more and more DS equations,
the solutions do not converge to the already known exact values of the Green's
functions.

Nevertheless, a remarkable feature of the unbiased approach is that for all five
theories studied in this paper, as we include more DS equations, the approximate
Green's functions actually {\it converge to limiting values}, and {\it these
limiting values are fairly accurate} -- several percent off from the exact
values for all of the Green's functions for all of the theories corresponding to
the possible pairs of Stokes sectors, as discussed above.

Finally, we have found a successful way to insert the missing information back
into the DS equations. Instead of using the unbiased {\it ansatz} of setting the
higher unknown Green's functions to zero, we replace $G_n$ by its {\it
asymptotic behavior for large $n$}. This procedure gives new and {\it extremely}
accurate numerical results for the Green's functions (many decimal places).
However, it does not eliminate all of the spurious theories associated with
different pairs of Stokes sectors; this can only be done by imposing
external additional conditions on the DS equations such as spectral positivity.
The use of the asymptotic behavior of $G_n$ for large $n$ suggests a new and
interesting general mathematical problem that has not been studied previously in
this context, namely, finding the asymptotic behavior of many-legged Green's
functions in higher-dimensional field theories.

One last remark: A simple way to force the DS equations to give sequences of
approximants that converge to the exact values of the Green's functions is to
require that the Green's functions all have formal weak-coupling expansions in
powers of a coupling constant. This approach has been known for a long time
\cite{R9}. To illustrate this idea we return to the trivial
differential-equation example above. We can demand that the solution to the
differential equation (\ref{e9.2}) be entire; that is, that the solution $f(x)$
have a convergent Taylor-series representation. This condition uniquely 
determines the unknown constant $C$ in (\ref{e9.3}): $C=0$. Unfortunately, it
does not recover the original function $f(x)$ in (\ref{e9.1}), which is
singular at the origin. Similarly, if we require that all Green's functions
have weak-coupling expansions, we immediately exclude the possibility of
using the DS equations to calculate Green's functions having nonperturbative
behavior. Indeed, if we ignore the possibility of nonperturbative behavior,
there is no reason to consider the DS equations at all because Feynman diagrams
give the perturbative representations of Green's functions.

\acknowledgments
We thank D.~Hook for assistance with some numerical calculations. CMB thanks the
Alexander von Humboldt and Simons Foundations, and the UK Engineering and
Physical Sciences Research Council for financial support.

\end{document}